\documentclass[fleqn,usenatbib]{mnras}

\usepackage{amsmath,amssymb} 

\usepackage{newtxtext,newtxmath,xcolor}
\usepackage[normalem]{ulem}

\usepackage[T1]{fontenc}

\DeclareRobustCommand{\VAN}[3]{#2}
\let\VANthebibliography\thebibliography
\def\thebibliography{\DeclareRobustCommand{\VAN}[3]{##3}\VANthebibliography}

\usepackage{graphicx}	
\usepackage{float}	
\usepackage{orcidlink}

\newcommand{\SNxx}[1]{\mbox{SN\,#1}}
\newcommand{\ATxx}[1]{\mbox{AT\,#1}}

\newcommand{\salt}{{\sc salt}}
\newcommand{\lmfit}{{\sc lmfit}}
\newcommand{\sncosmo}{{\sc sncosmo}}
\newcommand{\bg}{{\textit{nugent-sn91bg}}}

\newcommand{\msun}{M$_{\odot}$}

\newcommand{\PS}{\protect \hbox {Pan-STARRS}}

\newcommand{\sherlock}{\mbox{\texttt{sherlock}}}

\title[ATLAS100 transient survey]{ATLAS100 -- I. A volume-limited sample of supernovae and related transients within 100 Mpc}

\author[Srivastav et al.]{S. Srivastav$^{1}$\thanks{E-mail: shubham.srivastav@physics.ox.ac.uk}\orcidlink{0000-0003-4524-6883},
S. J. Smartt$^{1,2}$\orcidlink{0000-0002-8229-1731},
T. Moore$^{2,3}\orcidlink{0000-0001-8385-3727}$,
K. W. Smith$^{1,2}$\orcidlink{0000-0001-9535-3199},
D. R. Young$^{2}$\orcidlink{0000-0002-1229-2499},
M. D. Fulton$^{2}$\orcidlink{0000-0003-1916-0664},
\newauthor{
C. R. Angus$^{2}$\orcidlink{0000-0002-4269-7999},
M. Nicholl$^{2}$\orcidlink{0000-0002-2555-3192},
H. F. Stevance$^{1}$\orcidlink{0000-0002-0504-4323},
T.-W. Chen$^{4}$\orcidlink{0000-0002-1066-6098},
A. Pastorello$^{5}$\orcidlink{0000-0002-7259-4624},
J. Sommer$^{6}$\orcidlink{0000-0002-1154-8317},
}\newauthor{
F. Stoppa$^{1}$\orcidlink{0000-0002-3424-8528},
J. W. Tweddle$^{1}$\orcidlink{0009-0004-5681-545X},
J.~P. Anderson$^{7}$\orcidlink{0000-0003-0227-3451},
M. E. Huber$^{8}$\orcidlink{0000-0003-1059-9603},
A. Rest$^{3,9}$\orcidlink{0000-0002-4410-5387},
L. Rhodes$^{10,11}$\orcidlink{0000-0003-2705-4941},
}\newauthor{
L. J. Shingles$^{2,12}$\orcidlink{0000-0002-5738-1612},
A. Aamer$^{2,13}$\orcidlink{0000-0002-9085-8187},
A. Clocchiatti$^{14,15}$\orcidlink{0000-0003-3068-4258},
A. J. Cooper$^{1}$\orcidlink{0000-0002-4033-3139},
N. Erasmus$^{16,17}$\orcidlink{0000-0002-9986-3898},
}\newauthor{
J. H. Gillanders$^{1}$\orcidlink{0000-0002-8094-6108},
D. Magill$^{2}$\orcidlink{0009-0000-6521-8842},
G. Pignata$^{18}$\orcidlink{0000-0003-0006-0188},
P. Ramsden$^{2,19}$\orcidlink{0009-0009-2627-2884},
B. P. Schmidt$^{20}$\orcidlink{0000-0001-6589-1287},
X. Sheng$^{2}$\orcidlink{0000-0002-6527-1368},
}\newauthor{
J. G. Weston$^{2}$\orcidlink{0009-0002-9460-9900},
L. Denneau$^{8}$\orcidlink{0000-0002-7034-148X},
J. L. Tonry$^{8}$\orcidlink{0000-0003-2858-9657}
}
}
\date{Accepted XXX. Received YYY; in original form ZZZ}

\pubyear{2026}

\begin{document}
\graphicspath{{./}{figures/}}
\label{firstpage}
\pagerange{\pageref{firstpage}--\pageref{lastpage}}
\maketitle

\begin{abstract}

We present ATLAS100 -- a sample of 1729 supernovae and other explosive optical transients within $\sim 100$ Mpc observed by the ATLAS survey over a span of 5.75 years from 2017 September 21 to 2023 June 21. The volume-limited sample includes transients associated with galaxies with a spectroscopic redshift of $z \leq 0.025$, and spectroscopically classified transients within this redshift threshold where a host redshift was not available in existing catalogues. Our host galaxy list is constructed from aggregating all available galaxy redshift and distance catalogues. 
We carefully select all transients within a projected radius of 50\,kpc of these hosts. The ATLAS100 transient sample has a host galaxy redshift completeness fraction of $83$ per cent, consistent with expectations for the redshift completeness of local galaxy catalogues. Within this volume, the spectroscopic classifications are  87 per cent complete and we reclassify many ambiguous transients with joint light curve and spectroscopic considerations. Here, we release the catalogue together with compiled, binned and cleaned ATLAS photometry for all transients. We fit the light curve data to derive peak luminosities and characteristic timescales. We explore the sample characteristics, demographics and discuss the completeness and purity of the sample.  This is the first in a series of papers that will explore the rates and physical parameters of a complete and large sample of nearby supernovae and transients brighter than  $M \lesssim -16$. 

\end{abstract}




\section{Introduction}

The advent of modern robotic wide-field optical sky surveys, coupled with advances in detector technology and software techniques, has revolutionised the study of supernovae (SNe) and other transient phenomena. 
The All-Sky Automated Survey for Supernovae \citep[ASAS-SN;][]{2014ApJ...788...48S} has been capable of covering the entire visible sky rapidly at a sensitivity of $m\lesssim17.5\,$mag. Larger aperture facilities such as the Catalina Real-time Transient Survey \citep[CRTS;][]{2009ApJ...696..870D}, Palomar Transient Factory \citep[PTF;][]{2009PASP..121.1395L,2009PASP..121.1334R}, \PS1 \citep[PS1;][]{2010SPIE.7733E..0EK,2016arXiv161205560C}, the La Silla Quest \citep[LSQ;][]{2013PASP..125..683B} and the BlackGEM telescope array \citep{2024PASP..136k5003G} with footprints of $\sim 10$ square degrees, 
have surveyed sky areas of a few thousand square degrees with a cadence of a few days. The Asteroid Terrestrial-impact Last Alert System \citep[ATLAS;][]{2018PASP..130f4505T}, Zwicky Transient Facility \citep[ZTF;][]{2019PASP..131a8002B} and the Gravitational-wave Optical Transient Observer \citep[GOTO;][]{2022MNRAS.511.2405S} have extended the footprints to $\sim50-100$ square degrees, maintaining good sensitivity ($m\lesssim20$) with enhanced cadence all-sky capability. 
Additionally, dedicated spectroscopic facilities for classification and follow-up, such as the Public ESO Spectroscopic Survey for Transient Objects \citep[PESSTO and its extensions;][]{2015A&A...579A..40S}, the Spectral Energy Distribution Machine \citep[SEDM;][]{2018PASP..130c5003B} among others, have contributed immensely to the discovery and characterisation of novel and exotic transient subclasses \citep{Gal-Yam2017,modjaz_new_2019}, revealing new and hitherto uncharted regions in the duration-luminosity phase space \citep{2012PASA...29..482K}. 

Rapidly evolving transients (RETs) with characteristic timescales of $\lesssim 10$ days have been identified previously in surveys including \PS1 \citep{2014ApJ...794...23D}, PTF and Supernova Legacy Survey \citep{2016ApJ...819...35A}, Dark Energy Survey \citep{2018MNRAS.481..894P} and ZTF \citep{2023ApJ...949..120H}. RETs comprise a heterogeneous family of transients that include luminous fast blue optical transients \citep[LFBOTs;][]{2018ApJ...865L...3P}, the interaction-driven subclasses of Ibn and Icn SNe \citep{2022ApJ...926..125P}, the mysterious luminous fast coolers \citep[LFCs;][]{2023ApJ...954L..28N} found in elliptical host galaxies and the faint, fast and elusive class of kilonovae resulting from binary neutron star mergers \citep{2017ApJ...848L..12A}. The low luminosity region in the duration-luminosity diagram has been increasingly populated by a colourful variety of ``gap'' transients with peak luminosities intermediate to traditional SN classes and nova eruptions ($-10 \gtrsim M \gtrsim -16$). These include luminous red novae \citep[LRNe;][]{2017ApJ...834..107B}, intermediate luminosity red transients \citep[ILRTs;][]{2021A&A...654A.157C} and luminous blue variables \citep[LBVs;][]{2011MNRAS.415..773S}. The narrow line emission and optical faintness of these transients often mean their low signal-to-noise and low resolution spectra remain ambiguous. Calcium-strong transients \citep[CaSTs;][]{2010Natur.465..322P} comprise another family of faint and fast transients, with a preference for early-type hosts and remote locations \citep{2019ApJ...887..180S}. Type Iax SNe constitute a peculiar low-velocity and low-luminosity subclass of thermonuclear SNe Ia \citep{2013ApJ...767...57F} thought to arise from pure deflagrations \citep{2022A&A...658A.179L} in carbon-oxygen white dwarfs (WDs). Faint SNe Iax ($M \gtrsim -16$) occupy a similar region as other gap transients, with the most extreme observed events being up to 6 magnitudes fainter than normal SNe Ia at peak, synthesizing merely a few $10^{-3}$ \msun\ of radioactive $^{56}$Ni \citep[e.g.][]{2009AJ....138..376F,2009Natur.459..674V,2014A&A...561A.146S,2020ApJ...892L..24S,2021ApJ...921L...6K,2025ApJ...989L..33K,2026arXiv260209096Z}. These faint transients are only discovered in the nearby Universe, which means all-sky surveys are required to maximise the search volume. 

At the other extreme of the duration-luminosity phase space, super-luminous SNe \citep[SLSNe;][]{2011Natur.474..487Q} constitute the most energetic and powerful explosions in the Universe. Tidal disruption events \citep[TDEs; see][for a recent review]{2025arXiv251114911M} comprise a class of luminous nuclear transients (NTs) resulting from the disruption of a star that strays within the tidal radius of a massive black hole (MBH). TDEs emit across the electromagnetic spectrum from X-rays to radio \citep{2021ARA&A..59...21G}, and this emission can be used to independently decipher the properties and demographics of MBHs in otherwise dormant environments.

In addition to the discovery of new transient classes, recent surveys have also enhanced our understanding of the diversity within existing SN (sub)classes. These include the zoo of thermonuclear SNe \citep{Taubenberger2017}, stripped-envelope SNe (SESNe) with double-peaked light curves \citep[e.g.][]{2023A&A...678A.209K,2024ApJ...977L..41A,2024ApJ...966..199S} and interacting transients with precursor emission from months to years preceding explosion \citep[e.g.][]{2013ApJ...767....1P,2014ApJ...789..104O,2024ApJ...977..254D,2025arXiv250308768B}. These surveys also enable the study of large and homogeneous samples of SNe, vital for investigating the demographics, luminosity functions and volumetric rates for different SN types and subtypes. The Lick Observatory Supernova Search \citep[LOSS;][]{2011MNRAS.412.1441L} has been the benchmark for local Universe SN rates for over a decade, informing studies of progenitor channels
\citep{2011MNRAS.412.1522S}. This was based on observing targets from galaxy catalogues, followed by measured rates from wide-field surveys with no input galaxy bias. Determining SN rates in the local Universe is advantageous as the follow-up observational data (spectra and light curves) are of higher quality and inexpensive (in terms of telescope time) for the bright sources
compared to higher redshift surveys \citep{2017A&A...598A..50B,2021MNRAS.506.3330W}. The challenge is covering sufficient sky area to map the volume, which was achieved by ASAS-SN for different SN Ia subclasses \citep{2022ApJS..259...53C,2024MNRAS.530.5016D,2026arXiv260200223D} and SN II subclasses \citep{2025A&A...703A..34P}, and by PTF for volumetric rates of normal SNe Ia and CaSTs \citep{2018ApJ...858...50F,2019MNRAS.486.2308F}. The ZTF bright transient survey \citep[BTS;][]{2020ApJ...895...32F,2020ApJ...904...35P} provides a powerful magnitude-limited sample, leading to rates. The ZTF census of the local universe \citep[CLU;][]{2020ApJ...905...58D}, and now the ongoing complete astronomical transient survey \citep[CATS150;][]{2025TNSAN..64....1D} are volume-limited surveys aiming to classify all transients within $150$~Mpc. 

The ATLAS project was designed and funded by NASA to discover near-earth objects (NEOs) and potentially hazardous asteroids, or PHAs \citep{2011PASP..123...58T,2018PASP..130f4505T,2021PSJ.....2...12H}, through an all-sky survey system. The primary mission of mapping our neighbourhood in the solar system has been successful with the discovery of 1330 NEOs, 113 PHAs and 111 comets\footnote{\url{https://atlas.fallingstar.com}; as of 5 February 2026.}, including the potential impactor 2024 YR4 \citep{2025ApJ...984L..25B}
and only the third interstellar object, comet C/2025 N1 or 3I/ATLAS  \citep{2025arXiv250702757S}. 
The $1-2$ day, all-sky cadence is a rich resource for stellar astrophysics 
\citep{2018AJ....156..241H} and extragalactic transients. As described by \cite{2020PASP..132h5002S}, the ATLAS project continually processes the nightly data for extragalactic transients, releasing the discoveries immediately through the International Astronomical Union's Transient Name Server (TNS). With the discovery of \ATxx{2018cow}, ATLAS revealed the class of LFBOTs for the first time \citep{2018ATel11727....1S,2018ApJ...865L...3P}. ATLAS also discovered the first optical 
counterpart of an extragalactic fast X-ray transient (FXT) from the Einstein Probe
\citep{2024ApJ...969L..14G}, and the best candidate for a pair-instability supernova \citep[\SNxx{2018ibb};][]{2024A&A...683A.223S}. ATLAS routinely covers the skymaps of LIGO-Virgo-KAGRA gravitational wave detections to search for electromagnetic counterparts
\citep{2017ApJ...850..149S,2020A&A...643A.113A,2024MNRAS.528.2299S}. With over 23,000 SN candidates registered on the TNS and over 5,000 spectroscopically classified (as of January 2026), ATLAS is now a rich source for SN population studies \citep[e.g.][]{2022MNRAS.511.2708S,2025A&A...701A.128A,2026arXiv260203638E}, and provides open public access to light curves of any celestial source \citep{2021TNSAN...7....1S}.

Sampling the population of extragalactic transients within a fixed volume of the Universe provides an observational sample for direct comparison with theoretical stellar population and evolution studies. 
Biases that exist in magnitude-limited studies are mitigated but not removed completely. With the goal of sampling all transients within a fixed volume, in this paper we introduce ATLAS100, a volume-limited sample of transients within a redshift of $z \leq 0.025$ ($ D \lesssim 100$~Mpc) detected by the ATLAS survey over nearly six years of operation during 2017 September 21 to 2023 June 21. The sample features more than $1700$ transients, of which $\sim 40\%$ were discovered by ATLAS. The remaining 60\% were discovered and reported first to the TNS by other surveys but were detected independently by ATLAS. 
The full catalogue and calibrated, cleaned ATLAS light curves of all transients in the sample are publicly released with this paper. This is the first in a series of papers that will constrain the supernova population within this volume with high statistical significance, lack of host galaxy bias, having well measured light curves and good spectroscopic completeness.

\section{ATLAS100 sample definition}\label{sec:survey}

\subsection{ATLAS survey}
ATLAS was designed as an all-sky, high-cadence survey,   comprising multi-site, coordinated, wide-field telescopes, with two each in the northern and southern hemispheres. The project began  a 2-day cadence  northern sky survey (above $\delta=-50^{\circ}$) with the two units in Hawaii in January 2017.\footnote{The first ATLAS unit at Haleakala was operational from 2015, with Mauna Loa commissioned in 2017.} 
The addition of the two southern units in Sutherland (South Africa) and El Sauce (Chile) in 2022 brought all-sky coverage with a 1-day cadence (weather permitting at all sites).\footnote{In July 2025, a fifth ATLAS unit in Tenerife (Spain) with a different optical and detector design came in to operation. This unit was not used in the sample presented in this paper.}

As described in detail by \cite{2018PASP..130f4505T}, 
each unit has a 5.375$^{\circ}$ square detector at the focal plane of a ``Wright Schmidt'' telescope with an effective aperture of $0.5\,$m. This provides a $28.9$ square degree field of view, with typical $5\sigma$ limiting AB magnitudes of $18.5-19.5$ mag  for a $30\,$s exposure (depending on sky conditions). Images are obtained in two broadband filters -- cyan ($c$) and orange ($o$), roughly equivalent to a composite $g+r$ and $r+i$, respectively. The real-time data processing for stationary transients has been described by \cite{2020PASP..132h5002S}. Difference imaging is performed for the data from each ATLAS unit and all positive sources of 5$\sigma$ significance are catalogued and processed by the ATLAS transient science server. Typically, ATLAS observes a camera footprint 4 times within $40-60$ minutes, with the images separated by a temporal gap to optimise the successful linking of NEO detections and construction of tracklets. For stationary transients, we require at least 3 quality detections ($>5\sigma$ significance) within a $24\,$hr time period, defined by an integer whole MJD, to create an object that will then be assessed further. A convolutional neural network has been trained on human-verified real transients and known slow moving objects (point sources in 30\,s exposures) for real-bogus classification \citep{2020PASP..132h5002S}. Further refinements were made to the convolutional neural network described by \citet{2024RASTI...3..385W}, but these changes were made after the sample definition window.
In 2025, ATLAS introduced the Virtual Research Assistant (VRA) to significantly improve the selection of real, astrophysical transients \citep{stevance2025}, in an effort to reduce the workload on human scanners. The VRA is currently in operation as we continue the ATLAS100 survey, but did not play a role in the sample selection process for this paper. Access to the full history of  ATLAS photometry at any point on the sky is publicly available through the forced photometry server \citep{2021TNSAN...7....1S}, linked from the ATLAS home page.\footnote{\url{https://fallingstar-data.com/forcedphot/}}

\subsection{Sample definition}

\begin{figure}
\includegraphics[width=\linewidth]{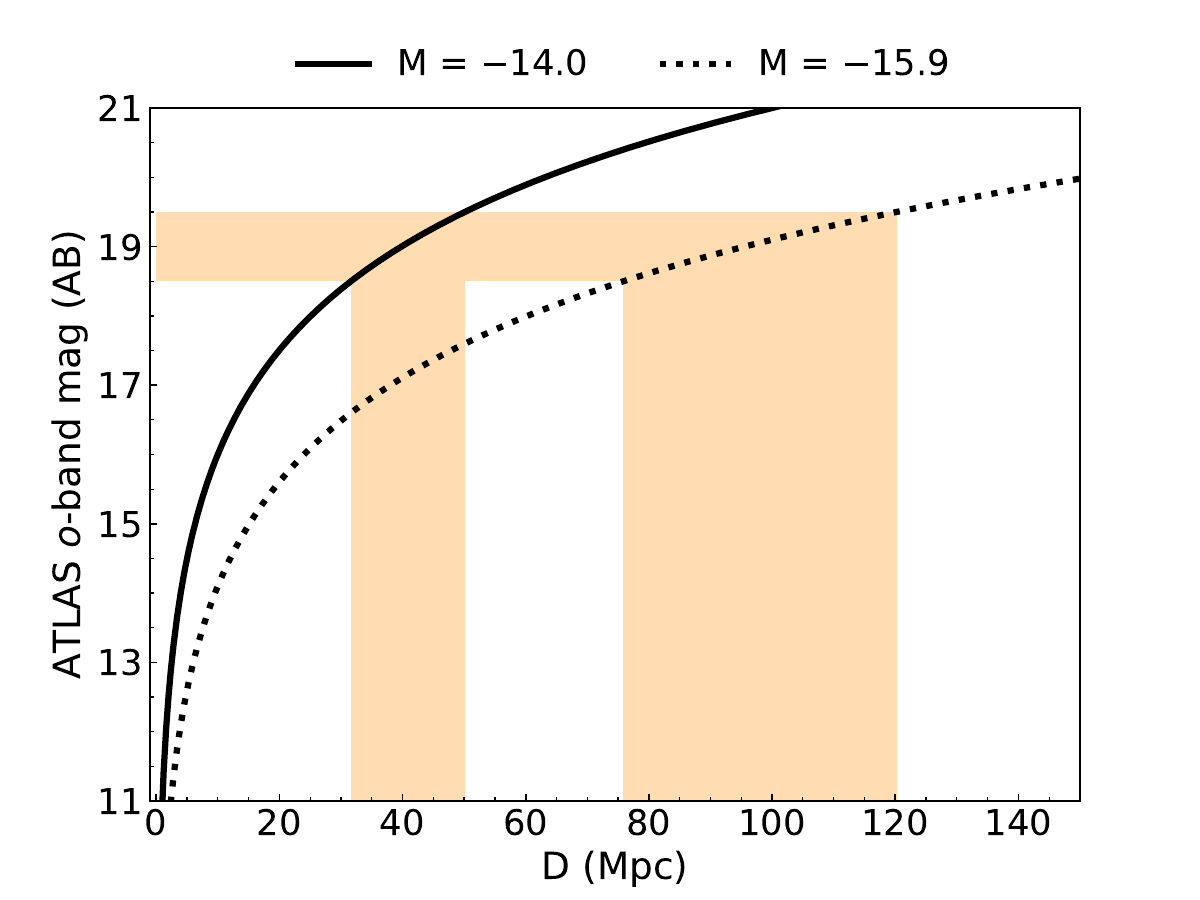}
\caption{The $o$-band magnitude versus distance for transients with absolute magnitudes $M_{o}=-14$ and $M_{o}=-15.9$. The solid orange regions project the sensitivity of $m_o=19.0\pm0.5$ to the distance to which a source would be detected. }
\label{fig:100MpcDM}
\end{figure}

The ATLAS100 sample aims to map a complete population of SNe and other extragalactic transients within a distance of $\lesssim 100$ Mpc. The time window for this sample covers a period of 5.75 years, spanning 2017 September 21 (MJD 58017) to 2023 June 21 (MJD 60116). 
The corresponding redshift for a luminosity distance of $D_{\rm L}\sim 100$\,Mpc depends on the assumed value of the Hubble constant $H_0$. We chose a redshift threshold of $z \leq 0.025$, which corresponds to a distance limit of  $D_{\rm L} \approx 109$\,Mpc 
assuming $H_0 = 70\,$km\,s$^{-1}$\,Mpc$^{-1}$ and a flat cosmology with $\Omega_{\rm M}=0.3$. 
This redshift/distance limit was chosen for several reasons. Within this volume, there is approximately 1 SN per day discovered, which over several years provides a significant statistical sample. The ATLAS sensitivity in the $o$-band (most frequently deployed) of 
$m_o \approx 19.0 \pm 0.5$ corresponds to $M_o \approx -16\pm0.5 - A_o$ at 100\,Mpc, where $A_o$ is the combined Milky Way (MW) and internal host galaxy extinction. For low values of extinction, this absolute magnitude will detect SNe Ia $\approx 3$ mag before peak at 100 Mpc (typically 2 weeks before maximum), and $M_o=-15.9$ is the peak measured magnitude of the kilonova \ATxx{2017gfo} \citep[average of $r$ and $i$ bands at +0.7 to +0.9 days post-merger;][]{2017PASA...34...69A,2017Natur.551...75S}. Figure\,\ref{fig:100MpcDM} illustrates the completeness limit for $m_o\lesssim19.0 \pm 0.5$ as a function of distance for $M_o=-15.9$ (appropriate for \ATxx{2017gfo}) and  $M_o=-14$, which corresponds to one of the faintest known SN Iax 2019gsc \citep{2020ApJ...892L..24S}. The absolute magnitudes of $M_o=-14$ and $-16$ also correspond to a normal SN Ia at 5 and 3 magnitudes before peak, at phases of roughly $-16$ and $-14$ days relative to maximum, respectively. 

The ability to find young, rapidly rising transients as opposed to those only within a few days of peak, at greater distances, was another motivation for choosing 100\,Mpc as the distance limit. We have been monitoring every transient ATLAS finds within 100\,Mpc in real time, due to the scientific promise of finding rare and intrinsically faint transients, or very young SNe. Since 2019 we have been alerting the community through AstroNotes for every transient within 100\,Mpc discovered by ATLAS with robust pre-discovery non-detections \citep[e.g.][]{2019TNSAN..17....1S} to encourage spectroscopic classification and further followup observations. We released 350+ AstroNotes from 2019 until June 2023. We also conducted programs to classify these nearby ATLAS transients at the 3.6\,m NTT, first through  the ePESSTO campaign 
\citep{2015A&A...579A..40S} and then through the ePESSTO+ consortium \citep{2023eas..conf...78I}, alongside 
access to the 2\,m Liverpool Telescope \citep[LT;][]{2004SPIE.5489..679S}. The PESSTO surveys classified 261 of the 1502 classified transients in the sample ($17.4\%$), behind only the ZTF group that accounted for 491 classifications ($32.7\%$).

Our initial selection of the ATLAS100 sample, both during the operational phase of the 2017--2023 survey and in retrospective checks, is based primarily on cross-matching transients with host-galaxy catalogues. This provides a secure host redshift, but we also supplement this with transients that lack a host redshift but have a spectroscopic redshift that places them within $z\leq0.025$. These were public spectroscopic classifications on the TNS, where the redshift of the transient (TNS object $z$) was within our threshold of 0.025. In such cases, the TNS object redshift is usually inferred from spectral template matching tools such as Superfit \citep{2005ApJ...634.1190H}, SuperNova IDentification \citep[\textsc{snid;}][]{2007ApJ...666.1024B}, \textsc{gelato} \citep{2008A&A...488..383H}, or \textsc{dash} \citep{2019ApJ...885...85M}. These redshifts are generally reported to two decimal places and are not as reliable as spectroscopic host redshifts, mostly due to the width of the spectral features and degeneracy between redshift, SN phase and ejecta expansion velocity. Occasionally, the classification spectra show narrow emission lines from the host, providing a secure redshift. A subset of this volume-limited sample was described by \citet{2022MNRAS.511.2708S}, where it was used to constrain the volumetric rates of SNe Iax. 

\begin{figure}
\includegraphics[width=\linewidth]{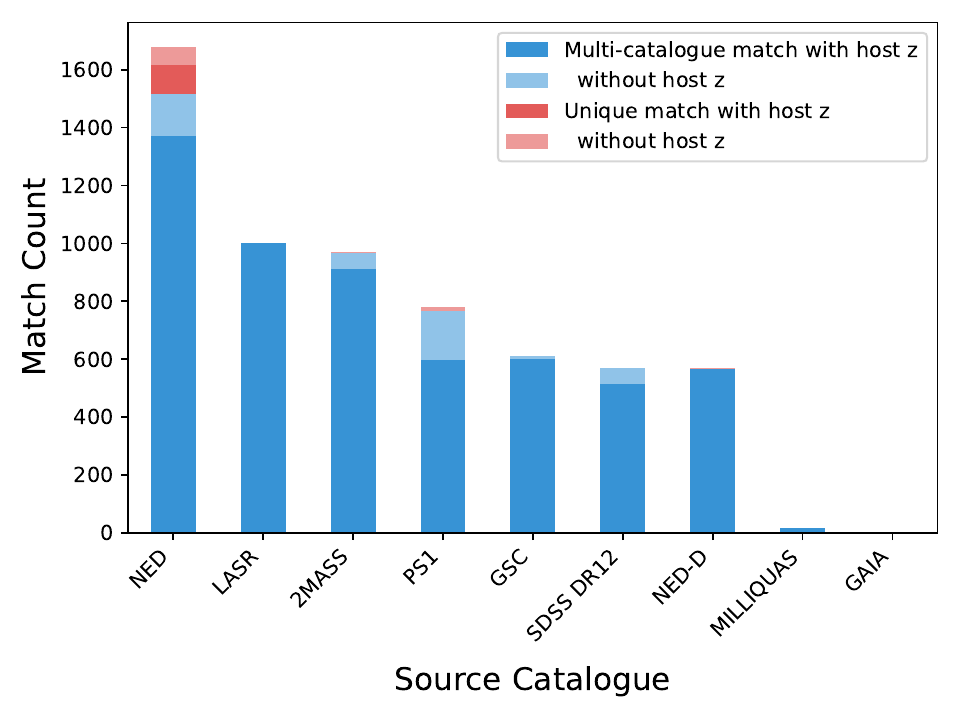}
\caption{Histogram showing the number of matches to different source catalogues mined by \sherlock\ for the host galaxies of 1729 transients in ATLAS100.}
\label{fig:sherlock_cat}
\end{figure}

We use the \sherlock \footnote{\url{https://github.com/thespacedoctor/sherlock}} \citep{Young_Sherlock_Contextual_classification_2023} Python package and its supporting backend database to assign a contextual classification to all candidate static transients in the ATLAS data stream. Sherlock achieves these classifications by mining an extensive library of archival catalogues of known astrophysical sources. These catalogues are hosted in a 4.5\,TB MySQL database, indexed by sky-location using the Hierarchical Triangular Mesh (HTM) indexing scheme. The catalogues include most of the major historical and ongoing astronomical surveys, including Gaia DR3 \citep{2023A&A...674A...1G}, SDSS DR12 \citep{2015ApJS..219...12A}, Pan-STARRS1 Science Consortium surveys \citep{2016arXiv161205560C,2020ApJS..251....7F,2020ApJS..251....6M} and 2MASS \citep{2006AJ....131.1163S}.  The \sherlock\ database also hosts many curated, source-specific catalogues, including the NED-D catalogue of galaxies and the Local AGN Survey \citep[LASr;][]{2020MNRAS.494.1784A} catalogue. Finally, \sherlock\ can dynamically query and cache data from the NASA Extragalactic Database (NED). Many of these catalogues report measured spectroscopic redshifts and/or redshift-independent distance measurements of the extragalactic sources they contain. Figure~\ref{fig:sherlock_cat} shows a histogram representing the major catalogues for the host galaxies in ATLAS100 cross-matched by \sherlock. NED and LASr serve as the primary sources for spectroscopic host galaxy redshifts for the transients in this sample.

\sherlock\ employs intelligent cross-matching of these catalogues and a decision-tree algorithm to predict the nature of each transient based on their cross-matched associations. The contextual classification returned by \sherlock\ is described in more detail by \citet{2020PASP..132h5002S}, and can include ``VS'' (variable star), ``AGN'' (if transient source is coincident with known and catalogued Active Galactic Nuclei or AGN), ``SN'' (likely to be a supernova based on association with a catalogued galaxy), and ``NT'' (nuclear transient, coincident with the core of a galaxy that is not a known AGN). During ATLAS survey operations, the contextual classifications from \sherlock\ are assigned automatically, and transients are promoted to a dedicated database table for manual review. These are the first targets that the on-duty team checks every day. 

For this summary paper, we compiled an initial list of all potential transients detected by ATLAS within $\sim 100\,$Mpc. This list includes transients identified by \sherlock\ as associated with a host galaxy with $z \leq 0.025$ or directly measured median distance $D \leq 100$ Mpc on NED, or with TNS Object $z \leq 0.025$. A projected radial distance of 50\,kpc was assumed by \sherlock\ as the maximum radius of association for galaxies with known redshifts. Transients that are more than 50\,kpc from their likely host have been identified 
\citep[e.g.][at $\geq150$\,kpc]{2011MNRAS.418..747M}, but these are rare and the hosts are not confidently associated with the SN. The most common subclass of transients with large offsets are CaSTs, which have been found between $10-45$\,kpc from their most likely hosts  
\citep{2014MNRAS.444.2157L}. 
Each object in this preliminary list was carefully examined manually, including the 
image stamps, \sherlock\ host association, the classification spectrum on TNS (if available) and if the reported redshift on TNS was consistent with the \sherlock\ host redshift (if available). Once a host galaxy association was established, we adopted the redshift of the identified host galaxy (from NED) as the redshift of the transient for further analysis. If a spectroscopic host redshift was not available, then the redshift derived from the classification report on TNS was adopted.

The preliminary list of transients compiled following the criteria defined above was subjected to a careful vetting process to identify and eliminate contaminants, described below. 


\subsection{Sample vetting and rejection of contaminants}\label{sec:vetting}

 The following categories of contaminants were identified and rejected in our vetting process:
 
\begin{enumerate}

\item \textit{Novae and nova candidates}. We do not include extragalactic novae in our sample, which is restricted to SNe and other extragalactic transients such as TDEs, CaSTs, and other gap transients. We therefore rejected confirmed novae with a spectroscopic classification on TNS, or identified as such through Astronomer's Telegrams (ATels). We also removed nova candidates, which we identify as unclassified transients in nearby host galaxies with a peak absolute magnitude $M \lesssim -9$ that were likely novae. At this absolute magnitude, novae are detected only in galaxies within $3-4$\,Mpc, with M31 and Centaurus A being good examples where we have uncovered multiple nova candidates. A total of nine candidate novae were identified and removed from the sample. We note, however, that we may have rejected unclassified, nearby and faint gap transients such as LBV eruptions and LRNe. Of the nova candidates that were removed, we highlight \ATxx{2022kjj} in M31, that shows a long-duration, undulating light curve. This transient does not look like a typical nova, and could be a gap transient in M31.

\item \textit{Cataclysmic Variables (CVs) and CV candidates}. Our preliminary extraction of $z \leq 0.025$ transients included several classified foreground CVs on the TNS with $z = 0$, these are common Galactic outbursts and were removed from the sample. A few unclassified candidates in our list were most likely foreground CVs, spatially coincident within the \sherlock\ crossmatch radius of a 100 Mpc galaxy. Telltale signs of these objects in the light curve are a blue colour at peak, a short rise time (typically a few days) followed by immediate decline, and in some cases evidence of recurrent activity in ATLAS, ZTF and/or PS1 history. A coincident, faint, blue stellar source was identified in PS1 reference images for some of these objects, corroborating their identification as foreground CVs. These were also removed from the sample. An example is \ATxx{2020krl} that was discovered by ATLAS with a projected separation of 18 kpc from the galaxy UGC 11262 at $z = 0.019$, but was later classified as a foreground CV \citep{2020TNSCR1561....1G}. 
    
\item \textit{Background SNe}. The projected radial offset of 50\,kpc was chosen to ensure we did not exclude SNe with large separations from their host (CCSNe and thermonuclear SNe can be found out to projected separations of $\sim 30\,$kpc and $\sim 50\,$kpc respectively, see Section~\ref{sec:hostSeparation}). However, a 50\,kpc radius circle projects to 1.7 arcmin at 100 Mpc, giving a sky search area of 9 square arcmin, which is large enough that background contamination occurs. The size of the search radius is a compromise between purity and completeness and for our initial selection sifting we set the radius to a large value within \sherlock.
For every transient, we visually inspected the multi-colour reference images  (Legacy Surveys, Pan-STARRS1, DSS, 2MASS, WISE) for any fainter host lying close to the transient that may be catalogued (with a photometric redshift) but lacks spectroscopic redshift or distance information. We flagged such transients as likely background contaminants, identified either during our day-to-day scanning process or during the final vetting stage for the preliminary sample. A total of 60 transients were identified during the survey period and removed as likely background sources. These are often confirmed as background SNe when a spectrum is obtained, but a majority of these are unclassified transients. 
For example, \ATxx{2020bbl} was crossmatched by \sherlock\ with the galaxy VV 457 ($z=0.005$) at a projected separation of 7.9\,kpc. However, the likely true host for this unclassified transient is a faint background galaxy (detected in Pan-STARRS 3$\pi$ survey with $m_{r_{\rm Kron}}=19.1$) that has a spectroscopic redshift, $z = 0.082$, from the Dark Energy Spectroscopic Instrument \citep[DESI;][]{2024AJ....168...58D}. \SNxx{2020dul} was similarly associated with NGC 5804 ($z=0.014$) with a projected separation of 20.2\,kpc, but was later classified as a background SN Ia at $z=0.07$ \citep{2020TNSCR.730....1S}. No obvious faint host is visible for \SNxx{2020dul}; it appears hostless, but the spectroscopic redshift is secure. 

There are therefore no background contaminants in the spectroscopically classified sample. We demonstrate in Section~\ref{sec:hostSeparation} that the offset distribution of the unclassified transients is comparable to the classified  subsample. It is possible that despite our careful vetting, a small number of the unclassified subsample are background contaminants within the 50\,kpc association radius of foreground $z \leq 0.025$ galaxies, and where the true host galaxy is too faint to be detected in archival images (see Section~\ref{sec:hostSeparation}).  Of the 60 transients that were rejected as likely background contaminants, we would expect no more than a few of these to have been misidentified and actually be transients lying within $z \leq 0.025$, based on both the high degree of spectroscopic completeness of local galaxy catalogues within $100-150$\,Mpc \citep{2018ApJ...860...22K}, and the unlikely combination of a faint transient occurring within a faint, $\leq 100$\,Mpc galaxy.

\item \textit{Refined redshifts from host galaxy emission lines}. 
For $\approx 83\%$ of the transients in the sample, the adopted redshift is derived from the catalogued host galaxy from \sherlock. Thus, only a small fraction ($\approx 17\%$) of transients in the sample lacked prior host galaxy redshift information and where the transient redshift was inferred from the classification spectrum. 
For a subset of these transients that were classified through our programs on LT or ePESSTO/ePESSTO+, 
we examined the two-dimensional (2D) spectra for signs of signal from the host galaxy. If host galaxy signal were present, we extracted the host spectrum and derived a more secure redshift from galaxy emission lines, primarily H$\alpha$, H$\beta$, and [O~{\sc iii}] $\lambda 5007$. In a few cases, the newly derived redshift from host emission features places the transient at $z > 0.025$; these were excluded from our sample. Examples include SN 2021adlt and SN 2021ubz. SN 2021adlt was classified by ePESSTO+ as a SN Ia at $z=0.021$ based on a \textsc{snid} comparison, however we derive a redshift of $z=0.027$ from host emission features. SN 2021ubz was classified by ePESSTO+ as a 91T-like SN Ia at $z=0.02$ using \textsc{snid}. However, the ATLAS light curve would peak at $M_o \approx -18.7$ at that redshift, faint for the over-luminous 91T-like Ia subclass. We derive a redshift of $z=0.030$ from the host galaxy emission lines instead, which implies a peak luminosity of $M_o \approx -19.6$, consistent with 91T-like SNe Ia. We note that this exercise was only carried out for the fraction of classification spectra where we have access to the raw 2D frames. These include 52 transients classified through either ePESSTO+ or our LT program, out of 294 where a prior host galaxy redshift was not available. We note that this method favours redshift determination for late-type hosts.
    
\item \textit{Obvious misclassification}. During our vetting process, we also checked whether the ATLAS light curve shape and peak luminosity was roughly consistent with the reported spectral type on TNS. This is pertinent in particular for the $17\%$ of transients that did not have prior host galaxy redshift information. During this process, we identified a handful of obvious erroneous classifications where the reported redshift is inconsistent with the light curve properties. An example is SN 2017ili, classified as a normal SN Ia by ePESSTO \citep{2017TNSCR1332....1L} at $z=0.025$. At that redshift, the ATLAS light curve has a peak luminosity of $M_o \sim -15.5$, too faint for a normal SN Ia. A careful analysis using \textsc{snid} shows good matches with SNe Ia at $z \sim 0.10 - 0.15$ instead, suggesting this is a background SN. Eight such identified events were also weeded out of the sample.
    
\end{enumerate}

\subsection{Spectral reclassification}\label{subsec:reclass}



During the vetting process, we examined the ATLAS light curve and the classification spectrum on TNS (where available) to ensure the reported spectral type and redshift (if derived from the classification spectrum) were broadly consistent with the light curve and luminosity. In a few cases, transients were classified through ATels but a classification report was not submitted to TNS. In other instances, the transient was classified on TNS but further follow-up observations (or published papers) conclusively show the transient belongs to a different spectral (sub)type. For example, some SNe IIP in ATLAS100 were reclassified as SNe IIb (or vice-versa) by detailed sample studies of SNe II observed by ATLAS \citep{2025A&A...701A.128A,2026arXiv260203638E}; these were incorporated in our sample.

Finally, during the vetting process we encountered classifications where the spectral features, light curve, redshift and/or host galaxy information was at odds with the reported spectral type on TNS. A full re-analysis of all available classification spectra was not performed for this study. Transients were reclassified only in the event of obvious misclassification, published ATels or papers, subsequent public light curve or spectroscopic information not available when the initial classification was reported to the TNS, or additional private follow-up data. 
The column labelled ``Spectral type'' in the catalogue file (released with this paper) reflects the adopted spectral type for each transient in the ATLAS100 sample. 
The catalogue data table includes a flag for the source of the adopted spectral type for each transient. For most of the sample, the adopted spectral type in ATLAS100 matches the spectral type reported to the TNS. In a few cases where our adopted spectral type is different, the flag specifies the source for the adopted ATLAS100 spectral type -- either from published literature or a reanalysis of publicly availably data, or private communication based on unpublished follow-up data. The ATLAS100 adopted spectral type is different from the TNS spectral type for 67 transients ($4.5\%$ of the classified subsample).


\subsection{Light curve data products}\label{subsec:lcbin}

We obtained ATLAS light curves for each SN in this sample using the publicly available \texttt{ATClean}\footnote{\url{https://github.com/srest2021/atclean}} package \citep{2023zndo...7897346R,2025ApJ...979..114R}. With \texttt{ATClean} we queried difference-image photometry from the ATLAS forced photometry (FP) server \citep{2021TNSAN...7....1S} at the TNS reported position of each SN within a time window ranging from $-500\,$d to $+1000\,$d since TNS discovery date (in the observer frame). The ATLAS FP data products were cleaned by \texttt{ATClean} following \citet{2025ApJ...979..114R}. We apply two quality control cuts to remove unreliable measurements using \texttt{ATClean} default settings. We make an uncertainty cut to remove light curve points with large uncertainties with a threshold of $\sigma_{\rm F}$ > 160 $\mu$Jy.  The ATLAS FP server returns the point-spread function fit reduced chi-squared statistic ($\chi_{\rm PSF}^2$) for each measured image. Any measurements with $\chi_{\rm PSF}^2$ > 10 were flagged as unreliable and removed. Finally, we bin the remaining measurements into nightly bins and adopt the 3$\sigma$-clipped average.

The light curve data release will generally include four data products for transients in ATLAS100. The data products include the cleaned and binned (1-day binning window) $o$-band and $c$-band data. For some transients, a baseline flux correction was applied to ensure the pre-explosion flux was consistent with zero. This offset was determined using a 200-day pre-explosion window, from $-220\,$d to $-20\,$d relative to epoch of TNS discovery (more details in Appendix~\ref{app:offset}). A baseline flux correction was applied for 275 light curves of 246 unique transients in ATLAS100 (out of 3449 light curves of 1729 unique transients). In case a correction was not required, the corrected data is a just a copy of the uncorrected ATClean output, but we include both data products for all transients for consistency. The light curve data is not corrected for any extinction along the line of sight. 

For example, for \SNxx{2019np}, the binned $o$-band data from ATClean is contained in the file named \texttt{2019np.o.1.00days.lc.txt}, while the baseline corrected data file is named \texttt{2019np.o.1.00days\_zp\_corr.lc.txt}. A similar naming structure is used for the $c$-band data. Table~\ref{tab:sample_LC} shows an abridged version of the binned $o$-band light curve for \SNxx{2019np}.

\begin{table}
\centering
\caption{An abridged sample of ATLAS light curve data from ATClean for one of the transients in ATLAS100. This table shows binned $o$-band light curve data for \SNxx{2019np}. 
The light curve data for all transients in ATLAS100 is publicly available as part of the ATLAS100 data release, the sample shown here is for guidance regarding its form and content.}
\begin{tabular}{lrrrrrr}
\hline
MJD & uJy & duJy & Nclip & Ngood & m & dm \\
\hline
58493.55 & 520.4 & 7.6 & 0 & 4 & 17.109 & 0.016 \\
58495.56 & 1393.5 & 6.2 & 2 & 8 & 16.040 & 0.005 \\
58497.54 & 2836.0 & 8.6 & 1 & 7 & 15.268 & 0.003 \\
58499.49 & 5112.0 & 30.0 & 0 & 1 & 14.629 & 0.006 \\
58501.48 & 7508.1 & 20.0 & 1 & 3 & 14.211 & 0.003 \\
\hline
\end{tabular}
\label{tab:sample_LC}
\end{table}

\subsection{Catalogue data products}\label{subsec:cat}

The catalogue data is also made publicly available with the ATLAS100 data release (see Data Availability). The catalogue data includes 40 columns; an abridged version is presented here for guidance (Table~\ref{tab:catalog}). The catalogue data includes the coordinates of the transient and its identified host galaxy, adopted redshift (from its host galaxy if available, from the TNS object $z$ if not), projected physical separation from the host, TNS discovery magnitude, and ATLAS peak magnitude.

\begin{table*}
\centering
\caption{An abridged sample of the catalogue data accompanying the ATLAS100 data release. The catalogue data includes 40 columns, only nine are shown here due to space constraints.}
\begin{tabular}{lrrrrrrrrr}
\hline
ATLAS Name & TNS Name & RA (deg) & Dec (deg) & Redshift & Spectral Type & Host Name & Angular Sep (\arcsec) & Physical Sep (kpc) \\
\hline
ATLAS17lyk & 2017hdz & 30.281734 & 31.889047 & 0.017312 & $-$ & NGC 0783 & 26.87 & 9.46 \\
ATLAS17koq & 2017gmr & 38.875728 & $-$9.354185 & 0.005037 & SN II & NGC 0988 & 34.87 & 2.7 \\
ATLAS17mal & 2017gvp & 355.444951 & $-$1.35088 & 0.022529 & SN Ia & UGC 12739 & 7.25 & 3.3 \\
ATLAS17jvk & 2017gpn & 54.438205 & 72.532953 & 0.007388 & SN IIb & NGC 1343 & 139.65 & 2.88 \\
ATLAS17nan & 2017hyf & 32.387518 & 21.249009 & 0.017132 & SN Ib & UGC 01652 & 8.85 & 3.08 \\
ATLAS18ebi & 2018ie & 163.504332 & $-$16.022679 & 0.014233 & SN Ic-BL & NGC 3456 & 37.33 & 8.63 \\
ATLAS18rso & 2018ctv & 21.466808 & $-$1.367085 & 0.018086 & SN Ia-91bg-like & NGC 0541 & 124.28 & 45.65 \\
ATLAS18mkz & 2017kam & 205.682987 & $-$24.335084 & 0.022496 & SN II-pec & ESO 509$-$ G 095 & 4.42 & 2.01 \\
ATLAS18bcmc & 2018aes & 207.074017 & 3.945747 & 0.003906 & ILRT & NGC 5300 & 31.69 & 3.12 \\
ATLAS18qgb & 2018bwo & 3.507145 & $-$23.193202 & 0.001558 & LRN & NGC 0045 & 50.82 & 1.67 \\

\hline
\end{tabular}
\label{tab:catalog}
\end{table*}

\section{Results and Discussion}\label{sec:results}

In this section, we present the general characteristics, statistics and demographics of the ATLAS100 sample. This is followed by a discussion on the sample completeness and purity, and the key science outputs and highlights (ongoing and imminent) from the ATLAS100 sample.

\subsection{Sample statistics and demographics}

\subsubsection{Redshifts and discovery magnitudes}

\begin{figure*}
\includegraphics[width=0.9\linewidth]{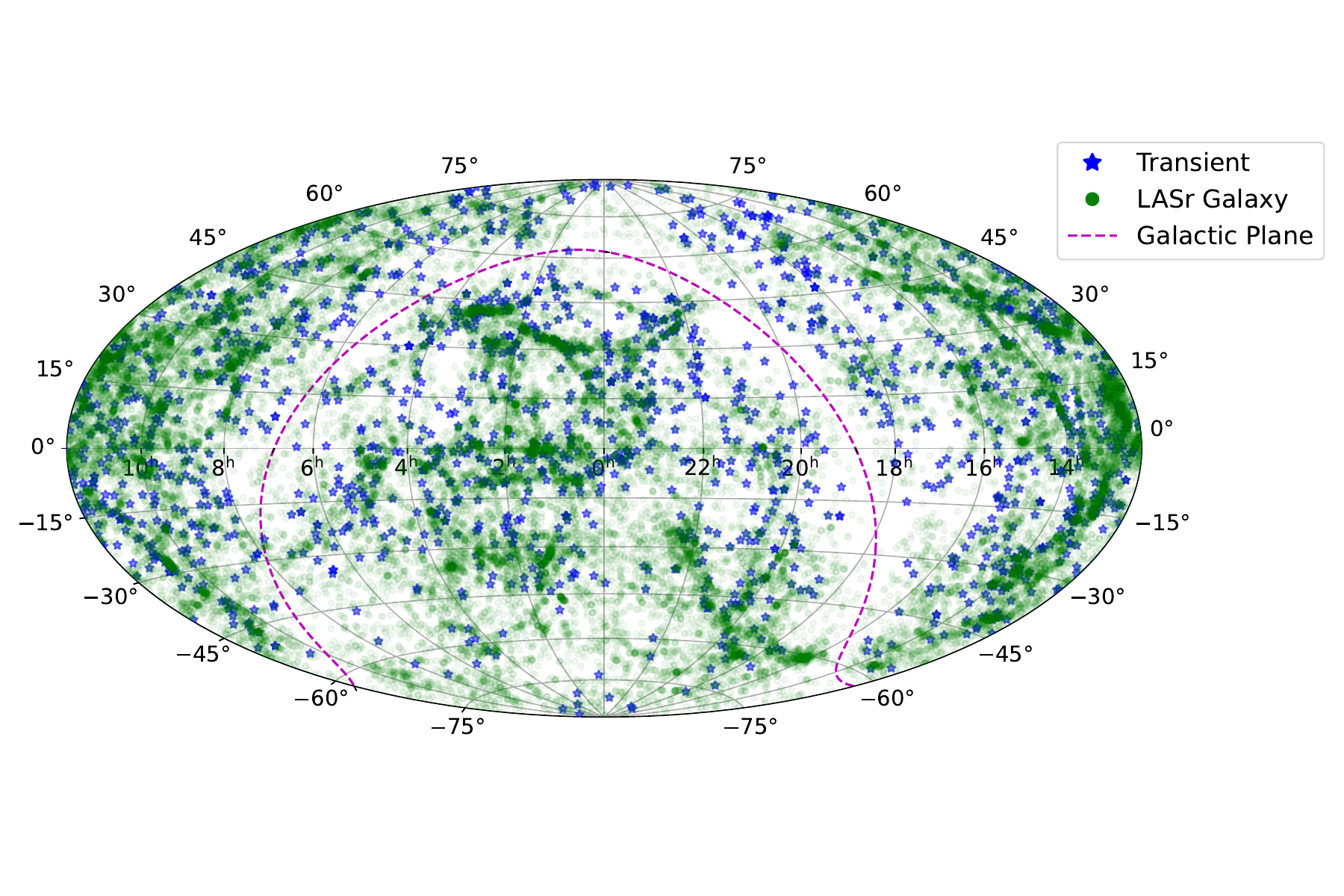}
\caption{Sky distribution of the ATLAS100 transient sample. Also shown (green circles) is the distribution of local galaxies within 100 Mpc from the Local AGN Survey \citep[LASr;][]{2020MNRAS.494.1784A}.}
\label{fig:SNdistribution}
\end{figure*}

The ATLAS100 sample consists of a total of 1729 transients that passed the vetting process described in Section~\ref{sec:vetting}.
Figure~\ref{fig:SNdistribution} shows the on-sky distribution of all the transients in the sample, generally tracing the distribution of nearby galaxies within 100 Mpc taken from LASr \citep{2020MNRAS.494.1784A}. Since the southern ATLAS units in Chile and South Africa have been operational only since late 2021, the number of transients in the sample south of declination $-50\degr$ (the limit of the northern Hawaiian units) are much fewer. There are also fewer transients in the Galactic plane due to high levels of foreground extinction.

\begin{figure*}
\includegraphics[width=0.7\linewidth]{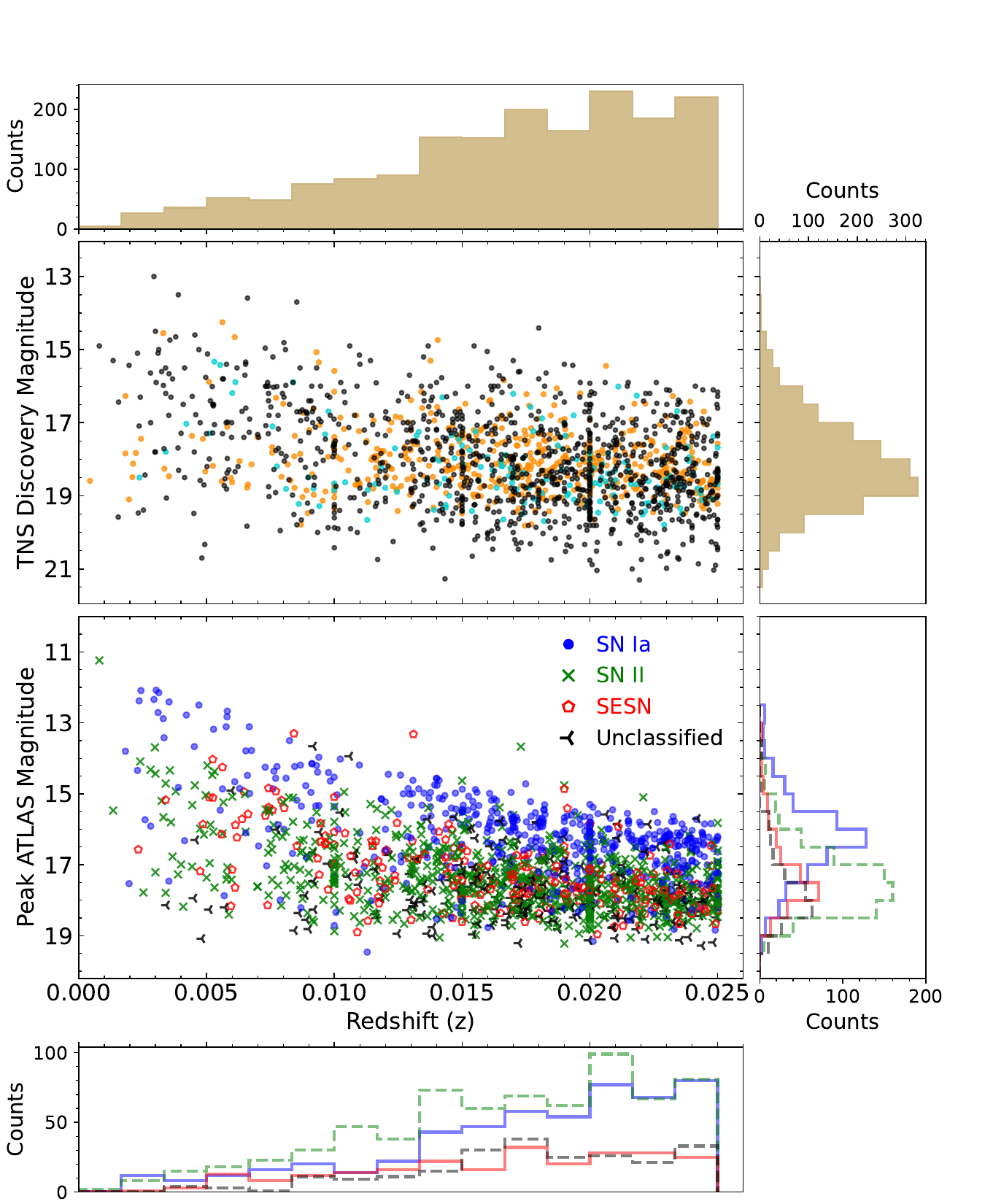}
\caption{Upper panel: distribution of TNS discovery magnitude versus redshift for all transients in ATLAS100. The ATLAS discoveries (689 of the total 1729) are shown in either orange or cyan, depending on whether the discovery filter was $o$-band or $c$-band. Non-ATLAS discoveries are shown as black circles. The histograms for redshift and discovery magnitude are for the full sample, with bin widths of $\Delta z = 0.00167$ (corresponding to $\sim 7\,$Mpc) and $\Delta\mathrm{mag} = 0.5$. Lower panel: distribution of peak observed ATLAS magnitude (without extinction correction) versus redshift, and the associated histograms for the key spectral types in the sample.}
\label{fig:discmag_z}
\end{figure*}

Figure~\ref{fig:discmag_z} shows the distribution of redshift and discovery magnitude (as reported to the TNS) for all transients in the ATLAS100 sample. The median redshift for the sample is 0.018 (0.006) and the median discovery magnitude reported to the TNS is 18.2 (1.2) mag, where the $1\sigma$ standard deviation is quoted in the brackets. We note that the discovery magnitudes are reported in different filters depending on the reporting survey (e.g. $o$ or $c$-band for ATLAS), and also sometimes reported in different photometric systems (e.g. Vega or AB). Nonetheless the plot is still useful, and suggests a median absolute magnitude of $-16.2$ (1.4) mag at discovery for a typical transient given the median discovery magnitude and redshift for the sample. ATLAS-discovered transients in the sample (689 out of 1729) are shown as either orange or cyan circles, depending on whether the discovery filter was $o$ or $c$-band. At redshifts of $z=0.010, 0.015, 0.020$, the effect of rounding the redshifts estimated from the spectra of the transients (lacking host galaxies) is visible as vertical overdensities. This indicates that even at $z\sim0.01$, galaxy redshift catalogues are not complete, with SNe signposting the underlying hosts which are catalogued in imaging surveys but without measured redshifts. 

\begin{figure}
\includegraphics[width=\linewidth]{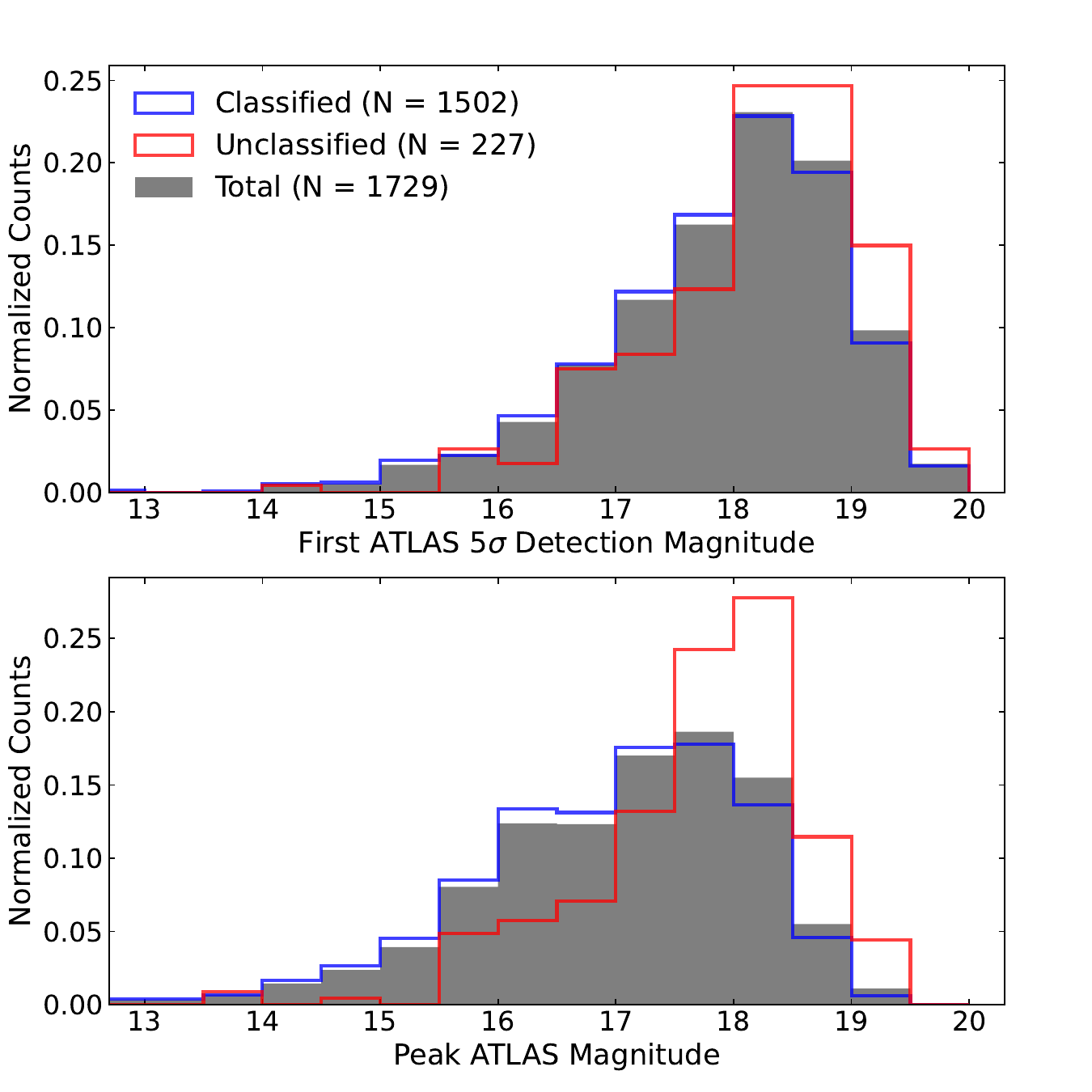}
\caption{Upper panel: histogram of first ATLAS $5\sigma$ detection for the classified (blue) and unclassified (red) sub-samples. The full sample is shown in grey, and the y-axis is normalized to the total number of counts for a direct comparison. Lower panel: ATLAS peak magnitude for the classified and unclassified sub-samples. The bin width is 0.5 mag.}
\label{fig:FirstMag}
\end{figure}

Figure~\ref{fig:FirstMag} shows the distribution of the first ATLAS $5\sigma$ detection magnitude (top panel) and the peak ATLAS magnitude (bottom panel) for the classified and unclassified subsets, along with the complete ATLAS100 sample. These are observed magnitudes, not corrected for foreground Galactic or host galaxy extinction. The subset of classified events ($87\%$ of the sample) is much larger, and thus a normalized histogram is shown for a meaningful comparison. Considering the sample (by definition) consists of the most nearby transients, a rather high fraction ($13\%$) is unclassified. It is therefore worth investigating if the subset of unclassified transients represents a distinct population. The median first ATLAS $5\sigma$ detection magnitudes are 18.06 (1.04) and 18.37 (0.93), for the classified and unclassified subsets, respectively. The median peak ATLAS magnitudes on the other hand are 17.12 (1.18) and 17.93 (0.94) for the classified and unclassified subsets, respectively. Although the difference is not significant, the peak ATLAS magnitudes suggest an overall fainter population for the unclassified events. The median redshift and median foreground Galactic extinction is identical for the two subsets, suggesting the difference in peak magnitude is not attributable to distance or MW extinction. We do not take host galaxy extinction into account here; however, we expect host galaxy extinction effects to average out given the large sample size. 

We also consider the possibility that the unclassified subset consists of a higher fraction of old SNe discovered after peak with declining light curves, emerging from solar conjunction. For old SNe discovered post maximum, the peak ATLAS magnitude (the same as the first ATLAS detection in this case) will be fainter than their true peak magnitude. These events are also less likely to be classified, since spectroscopic classification programs generally prioritise young SNe with robust pre-discovery constraints for the explosion epoch. To test this, we checked the fraction of events within the classified and unclassified subsamples that emerged from solar conjunction; i.e. with no pre-discovery non-detections within the last 30 days
in their ATLAS forced photometry light curve (described in Section~\ref{subsec:lcbin}). $78\%$ of classified events have pre-discovery non-detections whereas the fraction is $72\%$ for unclassified events, suggesting that old SNe are not significantly over-represented in the unclassified subset. This indicates that unclassified events in the ATLAS100 sample are broadly similar (albeit slightly fainter) to the population of classified SNe, rather than constituting a distinct subpopulation of faint and/or fast transients. The spatial offsets of the unclassified sample (from their identified host) tends to follow the classified sample as discussed in Section\,\ref{sec:hostSeparation}, further supporting our assertion that these are not a physically different population than the classified sample.

\subsubsection{SN types and subtypes}\label{subsubsec:SNtypes}

\begin{figure*}
\includegraphics[width=0.6\linewidth]{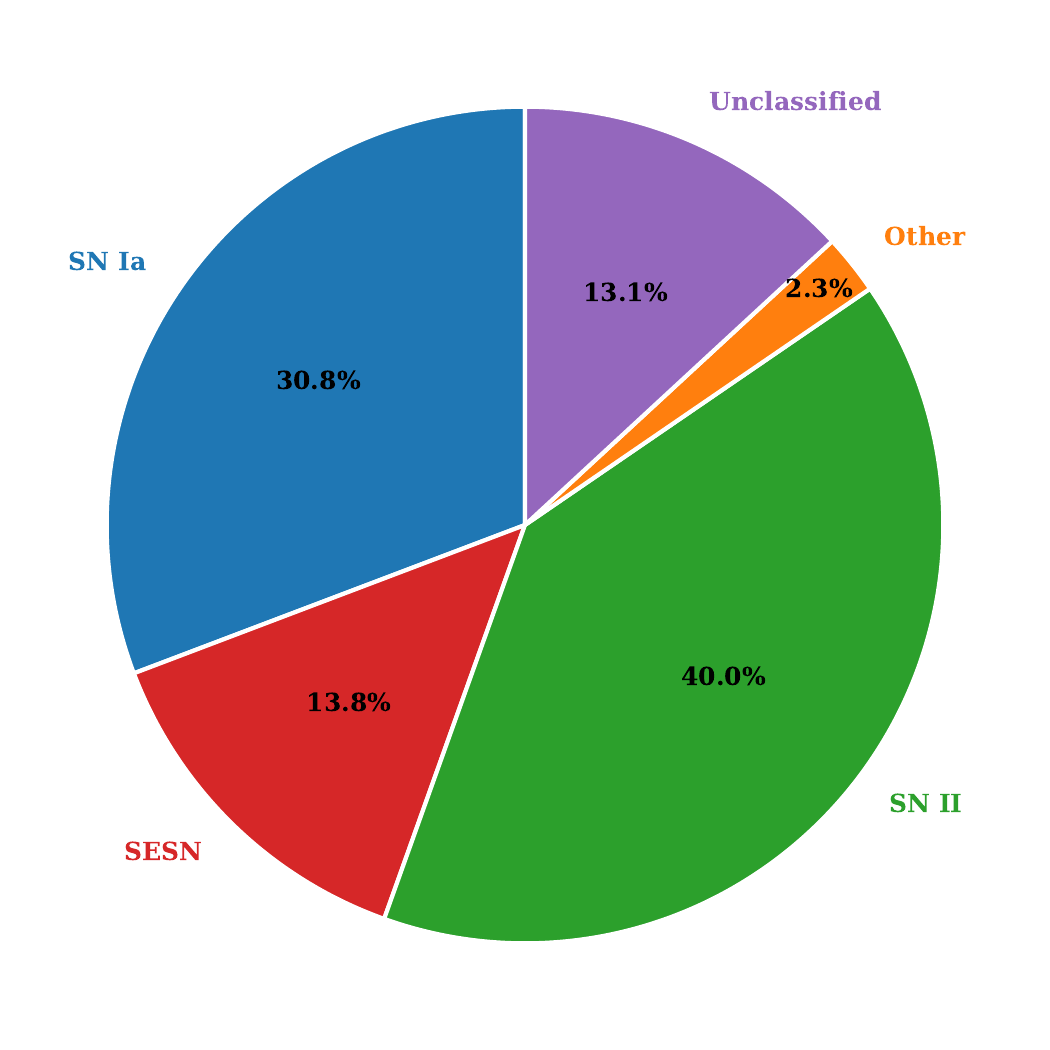}
\caption{Spectroscopic classifications for the transients in ATLAS100.}
\label{fig:PieChart}
\end{figure*}

Table\,\ref{tab:summary} lists a summary of the SN types and associated statistics and 
Figure~\ref{fig:PieChart} shows the observed fractional representation of the key spectral types in the ATLAS100 sample. We stress that these are observed fractions that are not corrected for the intrinsic light curve shape and luminosity of the transient, survey coverage and variations in sensitivity, foreground Galactic extinction, galactocentric distance from the host, etc. 
They represent the raw data set from which rates can be determined. 
A detailed analysis presenting corrected volumetric rates and luminosity functions for different SN types and subtypes will be presented in future work. 

Hydrogen-rich CCSNe or SNe II constitute the most abundant spectral type in the sample, comprising $40.0\%$ of the total sample and $46.1\%$ of the classified subsample. SESNe are relatively rare, comprising only $15.8\%$ of the classified events. CCSNe (SNe II + SESNe) account for $\sim 62\%$ of the classified subsample, whereas thermonuclear SNe Ia make up $35.4\%$. 
The fraction of SNe Ia (of all subtypes), at 35.4\% of the classified subsample, is significantly higher than that found by LOSS in their volume-limited sample 
\cite[24\%, within their 60\,Mpc volume after corrections for completeness;][]{2011MNRAS.412.1441L}. This is very likely due to the ATLAS completeness distance for the faintest CCSNe being lower than for the generally more luminous SNe Ia. This is evident in the histogram of redshift distribution as a function of spectral type (Figure~\ref{fig:discmag_z}, lower panel). SNe Ia in the sample have a median redshift of 0.019, whereas SNe II are generally closer with a median redshift of 0.017. We confirm this by constructing a sub-sample within 60 Mpc ($z \lesssim 0.014$). This ATLAS60 sample comprises 471 transients, of which 121 are SNe Ia ($25.7\%$), consistent with LOSS.

SNe II in the sample are classified into one of the following subtypes on TNS: II-norm, IIP, IIn and II-pec. Most of the objects classified as II-norm (just ``SN II'' on TNS) are in fact SNe IIP based on the typical plateau in their light curves. However, full light curve information is usually not available when classification reports are submitted to the TNS, especially when the spectrum is obtained within a few days of explosion, as is increasingly the case for these nearby SNe. The category of II-norm will include both IIP and IIL (plateau and linear light curves), which can be distinguished further with ATLAS light curve data. 
SNe IIn are relatively rare, constituting $6.1\%$ of the observed SN II population and $4.5\%$ of the overall CCSN population in the sample, the latter being consistent with the corrected SN IIn to CCSN ratios from PTF and ZTF BTS \citep{2023A&A...670A..48C}. Our observed SN IIn fraction is lower than the observed fraction in ZTF BTS by a factor of 2 \citep[$14.2\%$ of SNe II;][]{2020ApJ...904...35P}. Since SNe IIn are typically more luminous than normal SNe II, their observed fraction is higher in the magnitude-limited BTS sample. The ATLAS fraction of $6.1\%$ of all type II SNe is marginally lower than the $10.1\pm3\%$ from LOSS \citep{2011MNRAS.412.1441L} but within the uncertainties.  


There are 6 events that are classified as SN II-pec ($0.9\%$ of SNe II). Most of these are either confirmed or likely 1987A-like SNe II characterized by a distinct long rise to maximum and lack of the usual plateau feature, such as \SNxx{2021aatd} \citep{2024A&A...690A..17S}. 1987A-like events, thought to arise from blue supergiants, are rare and comprise a small fraction of SNe II \citep[$\approx 1.4\%$;][]{2023ApJ...959..142S}. Our raw rates are even lower but have not yet been corrected for completeness within the volume. 

The SESN subtypes present in our sample include all events with a root classification of SN IIb, Ib, SN Ic or SN Ib/c. This includes Ibn, Icn, Ib-pec, Ic-pec and broad-line Ic or Ic-BL. The SESN types represent those that are very likely the outcome of core-collapse in stars with $M_{\rm ZAMS}\gtrsim8$\msun.  We do not include the class of transients that are often labelled as SN Ib-Ca-rich into this category as they likely originate from a different explosion mechanism. 
Ca-rich or alternatively Ca-strong transients \citep[CaSTs;][]{2010Natur.465..322P} are placed into the ``Other'' category, since despite their spectroscopic resemblance to SNe Ib at early times, the nature of their progenitors remains an enigma. For a significant fraction, the host environments are more compatible with a thermonuclear origin rather than core collapse \citep{2019ApJ...887..180S}, although multiple progenitor channels have been suggested \citep{2020ApJ...905...58D}. 
This yields a total of 238 SESNe, accounting for $15.8\%$ of the classified subsample. 

SNe Ia are further subdivided into Ia-norm, Ia-91T, Ia-91bg, Ia-CSM, Ia-00cx, Ia-02es and Ia-03fg and Iax subtypes. An overwhelming majority ($88\%$) of SNe Ia in the sample belong to the Ia-norm category (i.e. classified as ``SN Ia'' on TNS). However, most of these classifications are directly from the TNS (with the exception of either egregious misclassification or objects with published papers in the literature and hence a confirmed membership of a particular subtype), and a more careful analysis of the spectra combined with light curve analysis will likely reveal a higher fraction of peculiar SN Ia subclasses in the sample \citep{2025A&A...694A...9B}. For instance, the ZTF SN Ia DR2 reported an observed fraction of $75\%$ for normal SNe Ia in their magnitude-limited sample \citep{2025A&A...694A..10D}, whilst also introducing a new subtype dubbed Ia-18byg. We defer a more rigorous analysis for SNe Ia in this sample, including sub-classification, computation of intrinsic volumetric rates and luminosity functions for the different Ia subtypes to a forthcoming paper. 

Of the 532 SNe Ia in the sample, 17 are classified as Ia-91T ($3.2\%$), 27 as Ia-91bg ($5.1\%$), 14 as Iax ($2.6\%$). There is a small number of events belonging to rare subclasses such as Ia-03fg (3), Ia-02es (2), Ia-CSM (1) and Ia-00cx (1). Here, we compare the observed fractions of Ia subtypes in ATLAS100 with the efficiency-corrected observed fractions from a volume-limited ($z < 0.06$) SN Ia sample from ZTF \citep{2025A&A...694A...9B,2025A&A...694A..10D}. The observed fraction of Ia-91T ($3.2\%$) in our sample is lower than the $12.2\%$ reported by ZTF \citep{2025A&A...694A..10D}. This is likely due to mistyping, suggesting a small fraction of SNe classified as Ia-norm on TNS belong to the over-luminous Ia-99aa or Ia-91T subtypes. The observed fractions of Ia-91bg, and the rare subtypes of Ia-02es, Ia-03fg and Ia-CSM in the sample are consistent with \cite{2025A&A...694A..10D}. The fraction of Ia-CSM is consistent with the volumetric rate of $\lesssim 0.2\%$ of the SN Ia rate \citep{2023ApJ...948...52S}.

SNe Iax are considered to be the most abundant peculiar subtype of SNe Ia \citep{2017hsn..book..375J}. Based on a subset of the sample presented in this work, \cite{2022MNRAS.511.2708S} estimated a volumetric rate of $15^{+17}_{-9}\%$ of the SN Ia rate. Furthermore, the Iax volumetric rate is expected to be dominated by faint events with $M_r \geq -16$ \citep{2022MNRAS.511.2708S}. The ATLAS100 sample is not complete at 100 Mpc for low-luminosity Iax events, with the faintest known object SN 2021fcg showing a peak luminosity of $M_r \sim -13$ \citep{2021ApJ...921L...6K}. It is thus unsurprising that the observed fraction of SNe Iax in the sample is lower than their intrinsic rates would suggest. 

CCSNe and SNe Ia, together with the unclassified events, make up $97.7\%$ of the sample. Finally, we compare our observed fractions of the key spectral types of SNe II, SESN and SNe Ia with an equivalent volume-limited version of the BTS sample. The latter is constructed by imposing $z \leq 0.025$ on the 5-yr BTS sample between 2018 May 01 and 2023 June 21. This volume-limited subsample of BTS contains 521 transients, all with a spectroscopic classification. The observed fractions in this BTS subsample are $43.0\%$ for SNe II, $32.2\%$ for SNe Ia and $16.9\%$ for SESNe, comparable to the observed fractions in ATLAS100 (Table~\ref{tab:summary}).

\begin{figure*}
\includegraphics[width=0.6\linewidth]{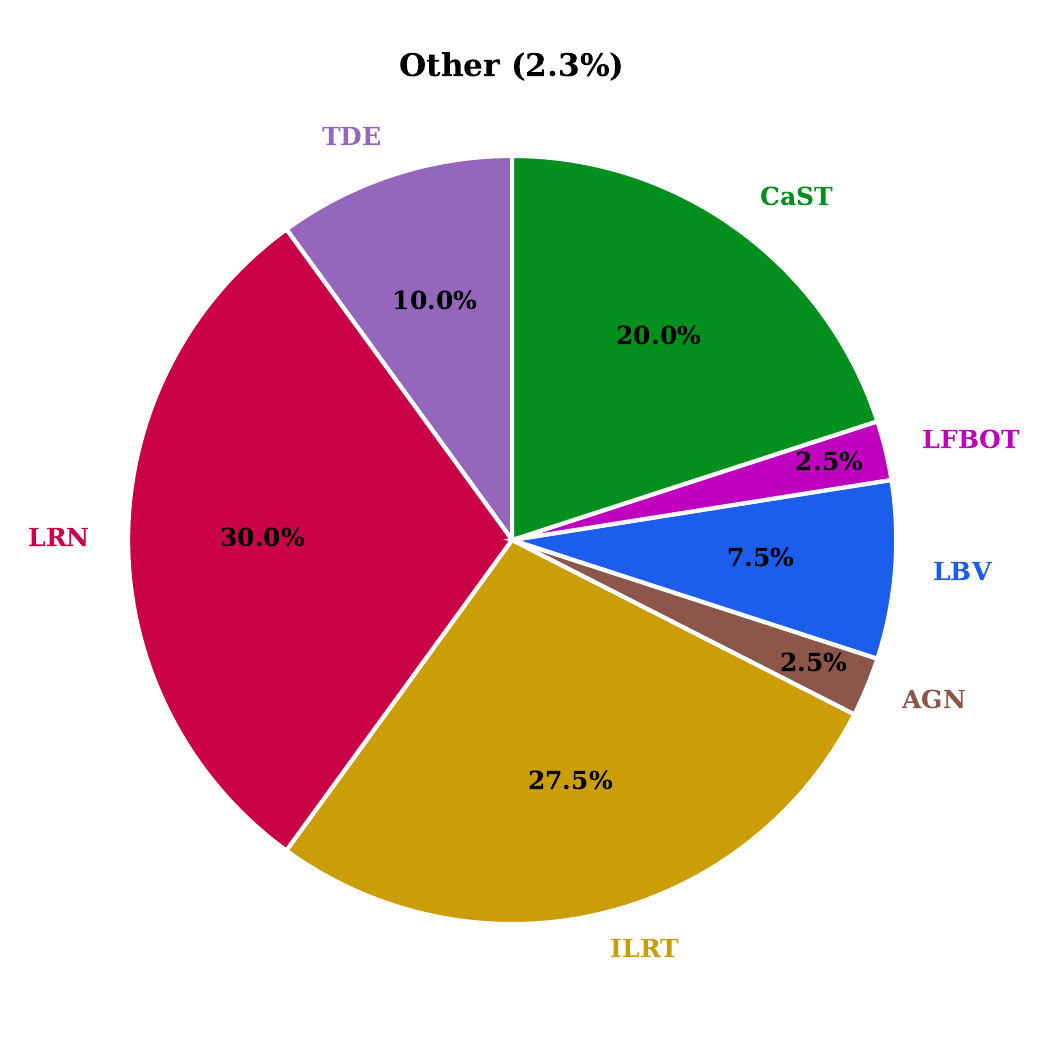}
\caption{Spectroscopic classifications for the subset of transients labelled as ``Other'' in Figure~\ref{fig:PieChart}.}
\label{fig:PieChartOther}
\end{figure*}

The remaining categories of transients are represented in the ``Other'' category -- these include 40 transients that include LBVs, LRNe, ILRTs, CaSTs and TDEs. 
Figure~\ref{fig:PieChartOther} shows the relative abundance of these rare transients in the ATLAS100 sample. Section~\ref{subsec:reclass} describes how transients in the sample were reclassified based either on reanalysis of public data, private follow-up data, or both. In particular, we reclassify a high fraction of gap transients (LBVs, ILRTs, LRNe) based on detailed follow-up data obtained for several individual events (private comm.). For example, our reclassification increases the number of ILRTs in ATLAS100 from 8 to 11, and the number of LRNe from 6 to 12. Most of these reclassifications were at the expense of transients classified as either LBVs, or as ``Impostor-SN'' on the TNS. The number of LBVs in the sample has diminished to 3 (from 8) after reclassification. In general, we note that LRNe and ILRTs are often misclassified as LBV eruptions. The 3 events classified on TNS as Impostor-SN are \SNxx{2019fxy}, \SNxx{2020bfw} and \SNxx{2022ryi}. Their light curves indicate intermediate peak luminosities ranging from $-14 < M_o < -10$, and the classification spectra show blue continuua with prominent Balmer emission features. Based on the light curve, we reclassify \SNxx{2019fxy} as a likely ILRT. \SNxx{2020bfw} exhibits a long-lasting eruptive phase, and prominent \ion{He}{i} emission in the classification spectrum \citep{2020TNSCR.402....1I}. Based on these characteristics, we reclassify \SNxx{2020bfw} as a LBV. \SNxx{2022ryi} is reclassified as a LRN based on extensive follow-up data (private comm.). After reclassification, LRNe ($30.0\%$), ILRTs ($27.5\%$) and CaSTs ($20.0\%$) constitute the bulk of these transients. It would appear that genuine giant outbursts from massive stars, often given the classification of LBV, are quite rare, and the bulk of these faint and intermediate luminosity gap transients are not from this type of massive star eruption. This requires further analysis of completeness within a smaller volume than 100 Mpc, and analysis of existing spectra and light curve data.  

Besides these subtypes, there are also 4 TDEs within this 100\,Mpc distance limit, implying a rate of roughly 1 optically bright TDE within 100 Mpc every 1.4 yrs. 
The archetypal LFBOT AT\,2018cow \citep{2018ApJ...865L...3P,2019MNRAS.484.1031P} is also part of the sample, which was detected across the electromagnetic spectrum from hard X-rays to radio 
\citep{2019ApJ...871...73H,2019ApJ...872...18M}. While other similar multi-wavelength emitting LFBOTs have been found \citep{2023ApJ...949..120H}, all are at significantly higher redshift indicating they are intrinsically rare by volume. Due to its uncertain physical nature, AT\,2018cow was originally classified as Ic-BL on the TNS (and still remains so as of February 2026). However it is now thought to be  most likely a disruption event rather than a supernova \citep{2023MNRAS.525.4042I}. 

There are 2 events in the sample classified as ``Galaxy'' on the TNS (\ATxx{2020wia} and \ATxx{2021aks}). Their ATLAS light curves resemble SNe (possibly faint or highly reddened) but the classification spectra are dominated by host galaxy light. 
\ATxx{2020wia} has a SN-like light curve and is coincident with the core of the galaxy 
2MASS J05485394+1911311, which has no catalogued redshift. A spectrum obtained by ePESSTO+ showed a red continuum and strong galaxy emission lines at $z=0.021$, but a definitive classification was not possible to determine from the spectrum \citep{2020TNSAN.207....1I}. 
\ATxx{2021aks} has a light curve that is also SN-like (albeit incomplete, with the early rise missing) and its spectrum was also dominated by galaxy light \citep{2021TNSAN..50....1T}. We reassigned these 2 transients as unclassified since their spectral type is unclear. 

\ATxx{2023dm} is an interesting nuclear transient and also in the Galactic plane ($b=3.23^{\circ}$) with a broad light curve ($\tau_{1/2}\sim50\,$days). It has a constant colour and is coincident with the core of a nucleated galaxy, both of which may indicate a TDE.   The ePESSTO+ classification spectrum shows narrow He\,{\sc i}, He\,{\sc ii}, and N\,{\sc iii} emission lines, possibly due to Bowen fluorescence at $z=0.021$. This event was classified as ``Other'' on the TNS \citep{2023TNSAN..10....1M}. It is not clear if this is nuclear activity in an AGN, but it has been at a consistently elevated level after the flare, which may also indicate a TDE plateau \citep{2024MNRAS.527.2452M}. This object is worth investigating further, but we refrain from classifying it as a TDE here, instead assigning it as ``AGN'' (Table~\ref{tab:summary}).

\subsection{Host galaxy separation}
\label{sec:hostSeparation}
\begin{figure*}
\includegraphics[width=0.75\linewidth]{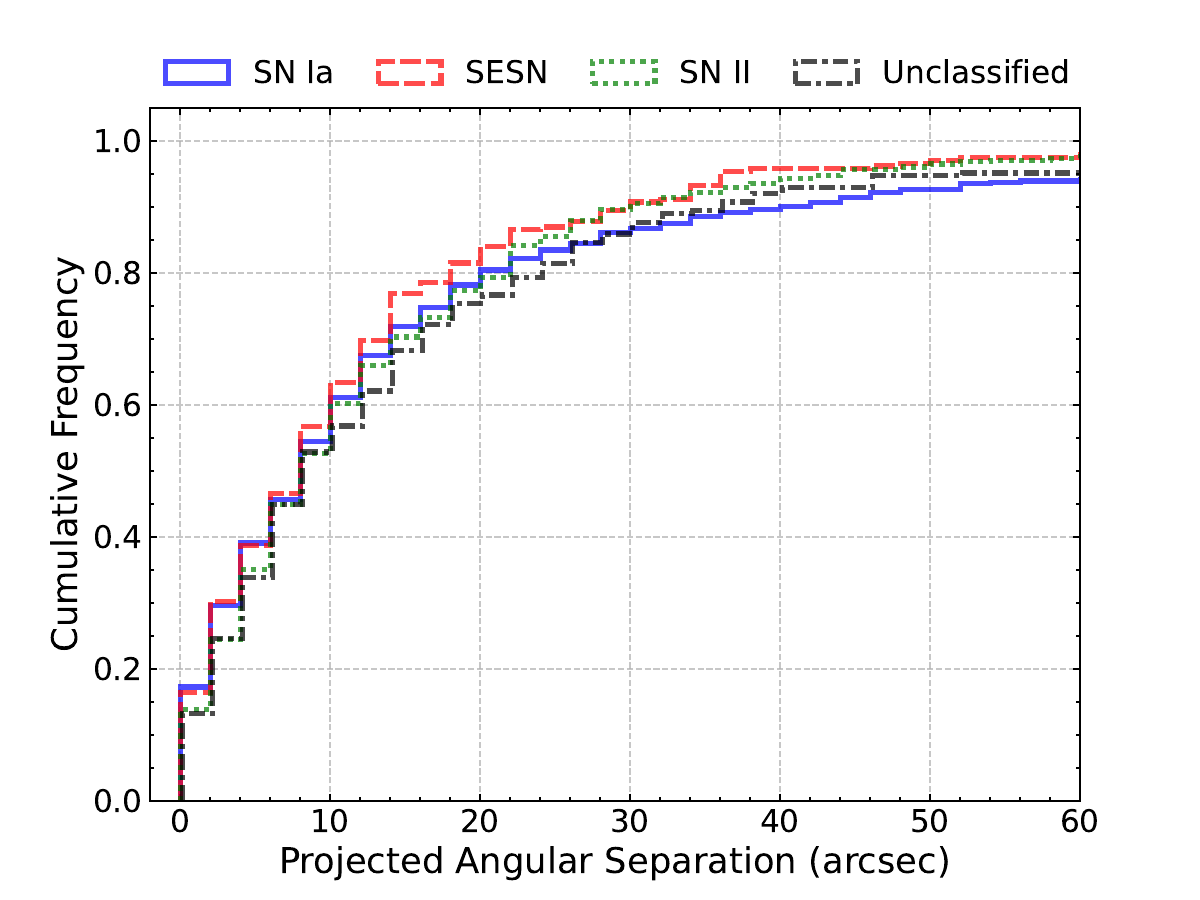}
\caption{Projected angular separation (\arcsec) from the host galaxy for the key distinct spectroscopic types in the sample: SNe Ia, SNe II, SESNe and unclassified transients.}
\label{fig:AngularSep}
\end{figure*}

\begin{figure*}
\includegraphics[width=0.75\linewidth]{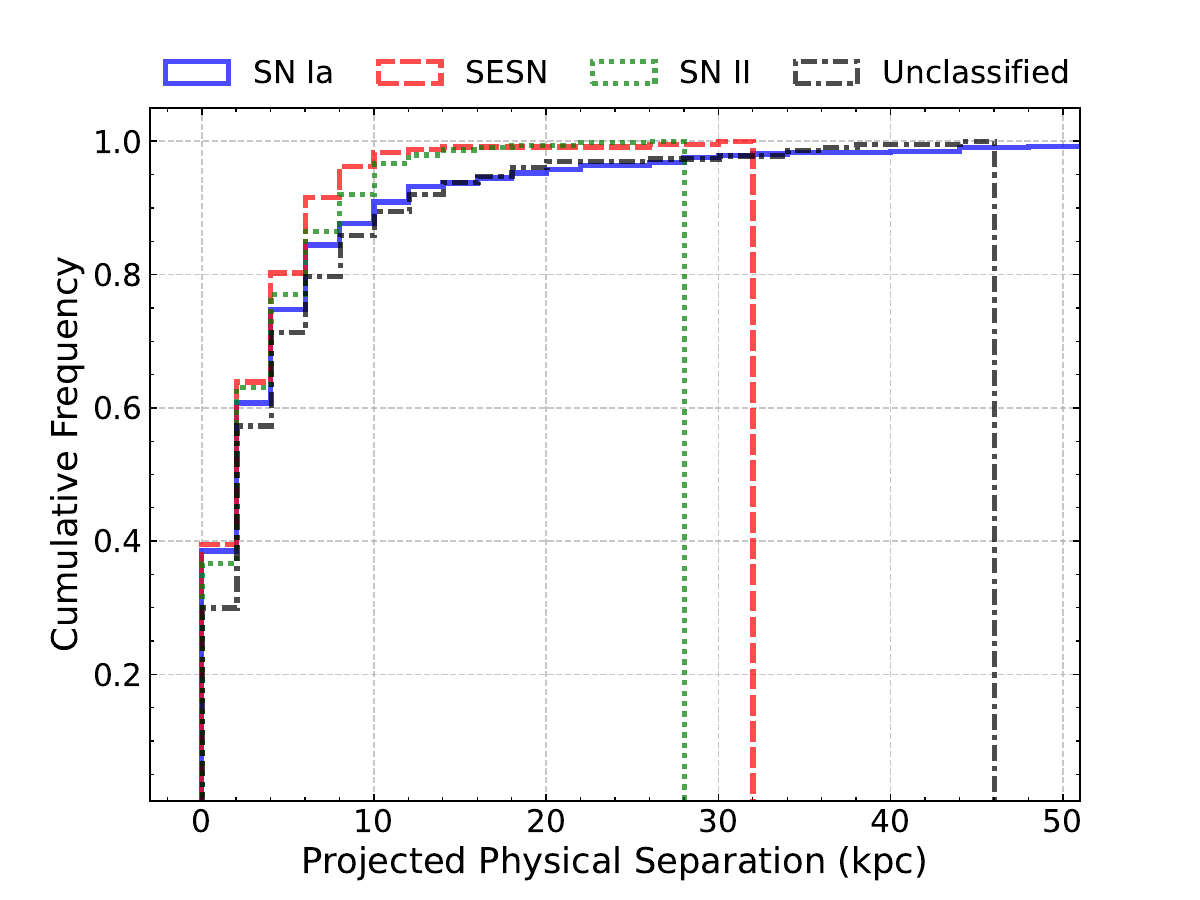}
\caption{Projected physical separation (in kiloparsec) from the host galaxy for the key distinct spectroscopic types in the sample, including SNe Ia, SNe II, SESNe and unclassified events.}
\label{fig:PhysicalSep}
\end{figure*}

Figures~\ref{fig:AngularSep} and \ref{fig:PhysicalSep} show  cumulative frequency diagrams of projected host galaxy separations for the key spectral types in the sample. We show them in both angular separation (arcsec) and in units of kiloparsec (kpc), assuming a distance from the host redshift. The histogram shows that the two CCSN families of SESNe and SNe II show a similar distribution for host galaxy separation, with all CCSNe occurring within $\sim 30$ kpc. Based on a literature sample of 177 CCSNe, \cite{2009MNRAS.399..559A} reported a strong deficit (excess) of SNe II (SESNe) in the central regions of their host galaxies. We do not recover this trend within the ATLAS100 sample, which consists of 688 SNe II and 240 SESNe (total of 928 CCSNe). The \cite{2009MNRAS.399..559A} sample was based on publicly announced SN discoveries from many different surveys, with differing sensitivities, and many were not difference image based. The larger ATLAS100 sample and uniform data and analysis methods does not recover a significant difference, although in each bin (up to a projected separation of 12\,kpc), the SESNe have a slightly higher frequency of occurrence. The SESN sample have a 90-percentile separation of 8\,kpc, compared to 10\,kpc for SNe II. A two-sample Kolmogorov-Smirnov (KS) test performed on the host galaxy separations for SESNe and SNe II yields a p-value of 0.4, suggesting there is no evidence these are drawn from different underlying populations. 

Like the CCSNe in the sample, the bulk of SNe Ia ($\sim 90\%$) occur within a projected separation of 12\,kpc. A small but measurable fraction of SNe Ia show an extended distribution; this is due to subclasses such as Ia-91bg showing a preference for remote locations in early-type galaxies \citep[e.g.][]{2025A&A...694A..10D}. There are 12 SNe Ia ($2.2\%$) that show projected separations of $R_{\rm proj}>30\,$kpc in our sample, 5 of which are classified as Ia-91bg. The distribution for unclassified events shows a more extended tail compared to CCSNe, but less so than SNe Ia. As suggested previously based on Figure~\ref{fig:FirstMag}, the majority of the unclassified subsample seems to be a combination of CCSNe and SNe Ia, rather than representing faint and/or unusual subclasses of transients. This further supports our assertion that the unclassified sample is not inherently different in nature to the classified sample, and does not introduce a strong and obvious bias into the relative rates. 

\begin{figure*}
\includegraphics[width=0.8\linewidth]{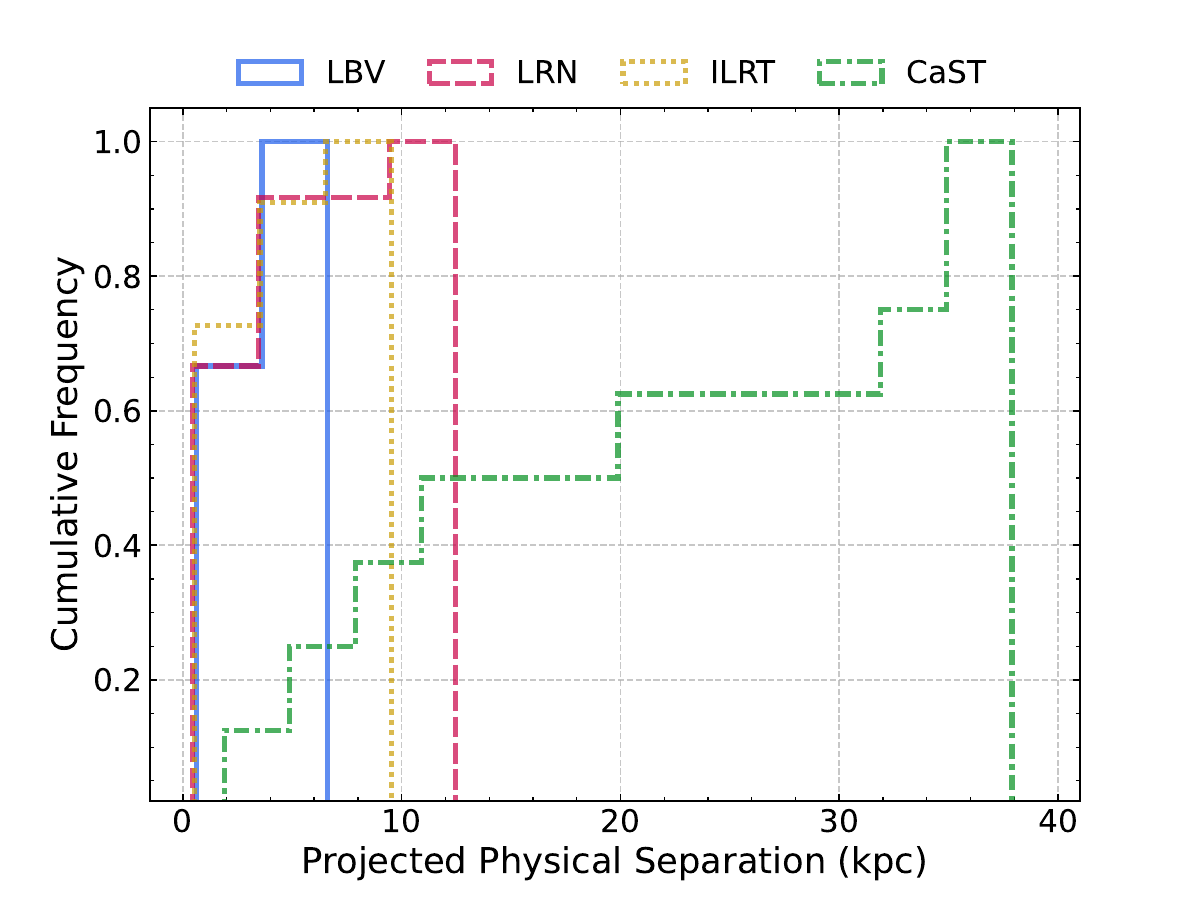}
\caption{Projected physical separation (in kiloparsec) from the host galaxy for gap transient families in the sample including LBVs, LRNe, ILRTs and CaSTs.}
\label{fig:PhysicalSepOther}
\end{figure*}

Figure~\ref{fig:PhysicalSepOther} shows a similar cumulative frequency histogram of projected radial separation (kpc) for gap transients in the sample -- 12 LRNe, 11 ILRTs, 8 CaSTs and 3 LBVs. LBVs, ILRTs and LRNe are confined to within $\sim 12$ kpc from their host galaxies. CaSTs on the other hand show a remarkably extended radial distribution. This is consistent with the already recognised tendency for these to occur in remote locations around early-type hosts \citep{2015MNRAS.452.2463F}, but our volume-limited sample demonstrates this quantitatively. 

\subsection{Light curve fitting}\label{subsec:lcfit}

\begin{figure*}
\includegraphics[width=\linewidth]{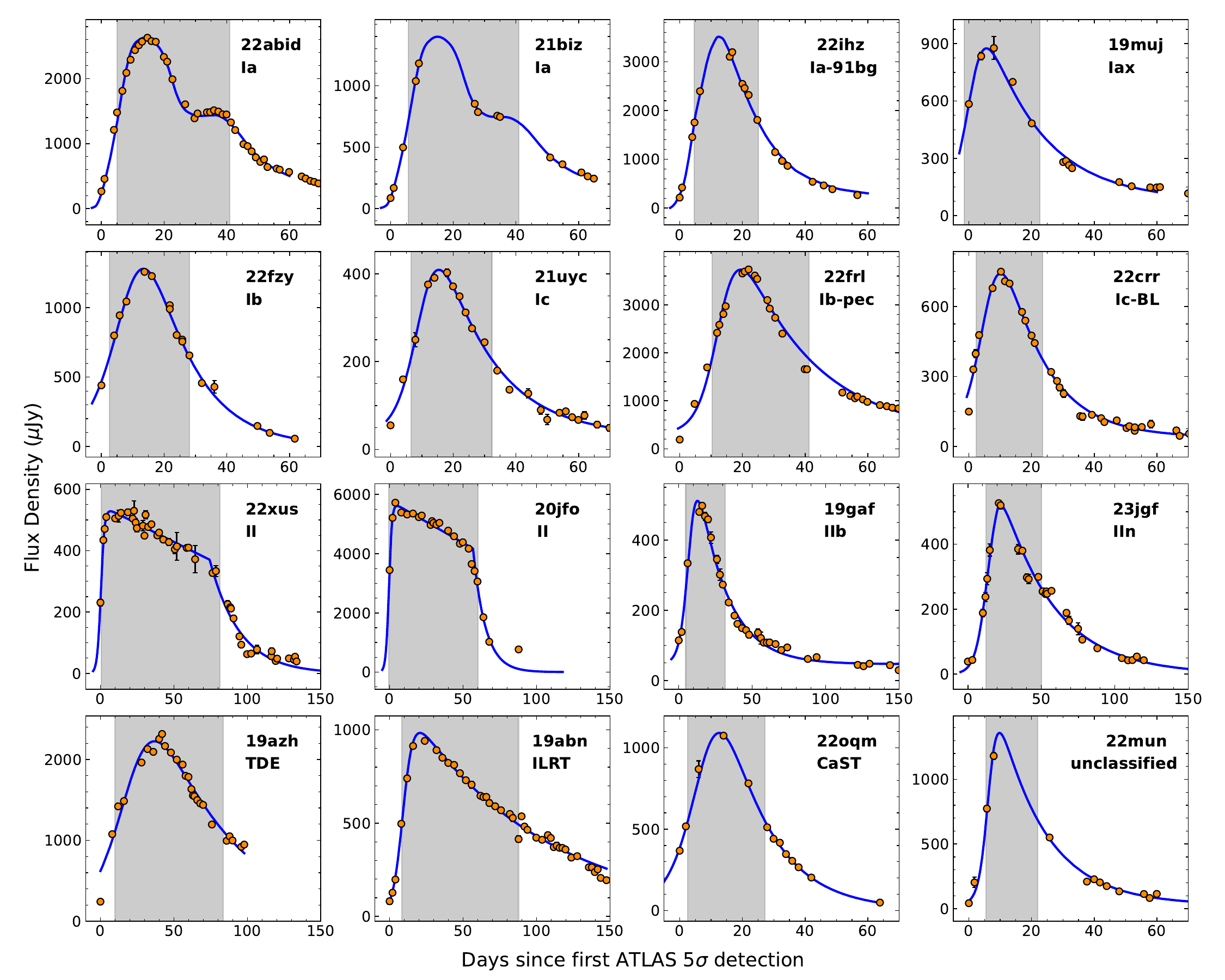}
\caption{Sample ATLAS $o$-band light curves of ATLAS100 transients across different spectral types and varying light curve quality and coverage. Also shown for each light curve is the best-fitting model that was used to derive the peak flux and the characteristic timescale or duration of the light curve. The shaded region represents the measured duration above half the peak flux.}
\label{fig:LCfits}
\end{figure*}

For all transients in the ATLAS100 sample, ATLAS photometry was binned and cleaned as described in Section~\ref{subsec:lcbin}. We fit the ATLAS light curves of all transients in the sample with a suitable model, discussed in more detail below, to estimate peak magnitudes and a characteristic timescale or duration for the light curve, defined as the time above half peak flux \citep{2020ApJ...904...35P}. In general, SNe Ia and Ib/c events show simpler light curves with a typical rise and fall, whereas SNe II tend to be longer lived with a more complex light curve morphology. The observing strategy of ATLAS is such that the $c$-band is only used around dark time. Most of the light curve data is thus obtained in the $o$-band and so we restrict light curve fitting to this band. 

We impose additional quality cuts in the light curve fitting routine. We require $\geq 2$ ATLAS non-detections in the time window between the TNS discovery epoch and 15 days prior. This condition ensures reasonable pre-explosion constraints and also eliminates transients emerging from solar conjunction and thus lacking a light curve peak and pre-discovery non-detections. Additionally, we require $\geq 2$ measurements during the rising phase of the light curve, $\geq 10$ points within TNS discovery epoch and 90 days after, and a peak observed flux, $f \geq 150\,\mu$Jy (corresponding to a peak observed magnitude of $\sim 18.5$ AB mag). Since the limiting magnitude of ATLAS is $\sim 19.5$, the latter ensures sufficient signal-to-noise (S/N) around peak for a robust light curve fit. These strict quality cuts removed 942 transients. A further 51 transients were rejected, either due to low S/N data, lack of photometric coverage around peak, or complex light curve morphology that was not fully captured by the simple analytical models. Our final sample consists of 736 transients that passed our quality cuts (Table~\ref{tab:summary}).The light curve fitting procedure for different SN types is described in more detail in the following subsections.

\subsubsection{SESNe}

We fit the light curves of SNe Ibc and SNe IIb in the sample that passed the quality cuts with the Bazin model \citep{2009A&A...499..653B}, an analytic model with an exponential rise and fall, defined as
\begin{equation}
f(t) = A \frac{e^{-(t-t_0)/\tau_{\rm fall}}}{1 + e^{(t-t_0)/\tau_{\rm rise}}} + B
\end{equation}
In general, with the exception of double-peaked light curves (e.g. shock-cooling tails in SNe IIb) or other peculiarities (such as precursor emission), we find the Bazin model provides an adequate fit to most SESN light curves. For SNe IIb, the early shock-cooling component (if present) was excluded from the light curve fit. The model has 5 free parameters ($A$, $B$, $t_0$, $\tau_{\rm rise}$ and $\tau_{\rm fall}$), and the fit was implemented using \lmfit\ \citep{lmfit}. The model constructed from the best-fitting parameters was used to derive the peak flux and the light curve duration; i.e. the time above half peak flux. 

\subsubsection{SNe Ia}

The SNe Ia that passed our quality cuts were also fit with the Bazin model described above. Although the Bazin model generally provides a reasonable fit for the rising light curves of SNe Ia, the post-peak light curve fit was found to be poor. Most SNe Ia show a pronounced secondary maximum or ``shoulder'' occurring roughly 3 weeks past the primary maximum \citep{2006ApJ...649..939K}, particularly conspicuous in the NIR but also noticeable in the redder optical bands. It is visible in well-sampled ATLAS light curves in the $o$-band. This feature is understandably not captured in the Bazin model and it affects the measurement of the light curve duration. Thus for SNe Ia, we instead use the \salt2 model \citep{2007A&A...466...11G} implemented in the \sncosmo\ package \citep{sncosmo}. We note that the secondary maximum is weak or absent in certain subclasses of SNe Ia such as Ia-91bg, Iax, Ia-02es and Ia-03fg. For Ia-91bg, we use the 
\bg\ model \citep{2002PASP..114..803N} in \sncosmo. The Bazin model produces a better fit for the other peculiar subclasses Iax, Ia-02es and Ia-03fg, and we thus use the derived peak fluxes and durations for these from the Bazin fit.

\subsubsection{SNe II}

The SN II light curves were fit with the analytical model presented by \cite{2019ApJ...884...83V}. This model is similar to the Bazin with two additional free parameters for the plateau component often seen in SN II light curves. SNe II are a diverse family of transients with a rich variety in light curve morphology, and in some cases this simple model is unable to capture light curve features, in which case the estimated peak and light curve duration is unreliable. The fits were inspected visually and discarded if the overall light curve shape and/or peak was poorly fit.

\subsubsection{Unclassified and Other Transients}

For unclassified transients, and transients belonging to the ``Other'' category (Figure~\ref{fig:PieChartOther}), we use the Bazin model to fit the light curves. The Bazin model overall does reasonably for unclassified SNe, although some are likely SNe II. The Bazin model also produces reasonable fits for ILRTs, CaSTs and TDEs. In case the light curve fits were found to be inadequate, and if the light curve coverage/cadence was good, the duration ($\tau_{1/2}$) was estimated simply from the observed light curve.

\subsection{Duration-luminosity relations and science highlights  }\label{subsec:science}

The peak fluxes derived from the model fits were corrected for foreground MW extinction. For each transient, we also derive a peak observed flux and observed duration, measured directly from the binned, cleaned ATLAS light curve. For well-sampled light curves, the observed values show a good match with the model-derived parameters. However, when there are gaps in the light curve coverage, the directly measured values are unreliable and generally underestimate the light curve duration and/or the peak flux. 
Figure~\ref{fig:LCfits} shows ATLAS $o$-band light curves for selected events belonging to different spectroscopic types, along with the best-fitting model. The shaded region represents the time above half the peak flux derived from the model. Table~\ref{tab:summary} summarises the the different spectroscopic subclasses in ATLAS100 and their median durations and peak luminosities derived from the light curve fitting.

\begin{table}
\centering
\caption{Summary of the statistics, peak luminosities and characteristic timescales for the different spectral types in the ATLAS100 sample. $N_{\rm tot}$ is the total number of transients for a given spectral type and  $f_{\rm tot}$ is the percentage of the 1502 total classified events. $N_{\rm cut}$ is the number of transients that passed the stringent light curve quality cuts (defined in Section~\ref{subsec:lcfit}) that were used to estimate the peak luminosity and duration. The median peak $o$-band magnitude and median duration is shown for all spectral subtypes with $N_{\rm cut} \geq 3$. The $1\sigma$ standard deviation in the peak absolute magnitude and duration is indicated in the brackets.}
\begin{tabular}{rrrrrr}
\hline
Spectral  & $N_{\rm tot}$ & $f_{\rm tot}$ & $N_{\rm cut}$ &  Peak Mag &  Duration \\
Type &  & (\%) &  & ($o$-band) & (days)\\
\hline
{\bf Type II} & {\bf 692} & {\bf 46.0} & {\bf 272} \\
II-norm & 643 & 42.8 & 251 & $-17.07$ (0.74) & 83.3 (28.7) \\
IIn & 42 & 2.8 & 19 & $-17.81$ (0.95) & 64.8 (26.6) \\
II-pec & 6 & 0.4 & 2 & & \\
\\
{\bf Type Ia} & {\bf 532} & {\bf 35.4} & {\bf 287} \\
Ia-norm & 467 & 31.1 & 248 & $-18.65$ (0.55) & 33.5 (5.5) \\
Ia-91bg & 27 & 1.8 & 16 & $-17.65$ (0.77) & 20.6 (0.2) \\
Ia-91T & 17 & 1.1 & 9 & $-19.01$ (0.40) & 40.5 (1.1) \\
Iax & 14 & 0.9 & 9 & $-16.62$ (1.29) & 29.2 (3.1) \\
Ia-03fg & 3 & 0.2 & 2 & & \\
Ia-02es & 2 & 0.1 & 2 & & \\
Ia-CSM & 1 & 0.1 & 0 & & \\
Ia-00cx & 1 & 0.1 & 1 & & \\
\\
{\bf SESN} & {\bf 238} & {\bf 15.8} & {\bf 122} \\
Ic & 77 & 5.1 & 37 & $-17.18$ (0.75) & 29.3 (11.6) \\
Ib & 72 & 4.8 & 44 & $-17.12$ (0.43) & 30.7 (6.3) \\
IIb & 44 & 2.9 & 22 & $-17.27$ (0.84) & 28.7 (13.7) \\
Ib/c & 16 & 1.1 & 6 & $-17.47$ (0.25) & 25.6 (3.0) \\
Ic-BL & 13 & 0.9 & 4 & $-17.22$ (0.65) & 17.4 (6.8) \\
Ib/c-pec & 9 & 0.6 & 5 & $-17.47$ (0.15) & 31.6 (7.1) \\
Ibn & 7 & 0.5 & 4 & $-17.56$ (0.30) & 13.3 (2.6) \\
Icn & 1 & 0.1 & 0 & & \\
\\
{\bf Other} & {\bf 40} & {\bf 2.7} & {\bf 13} \\
LRN & 12 & 0.8 & 2 & & \\
ILRT & 11 & 0.7 & 5 & $-13.43$ (0.83) & 57.2 (18.7) \\
CaST & 8 & 0.5 & 3 & $-16.30$ (0.52) & 24.8 (3.4) \\
TDE & 4 & 0.3 & 2 & & \\
LBV & 3 & 0.2 & 1 & & \\
AGN & 1 & 0.1 & 0 & & \\
LFBOT & 1 & 0.1 & 0 & & \\
\\
{\bf Unclassified} & {\bf 227} &  & {\bf 42} & $-17.48$ (1.22) & 32.0 (32.6) \\
\\
{\bf Total} & {\bf 1729} &  & {\bf 736} \\
\hline
\end{tabular}
\label{tab:summary}
\end{table}

\begin{figure*}
\includegraphics[width=0.9\linewidth]{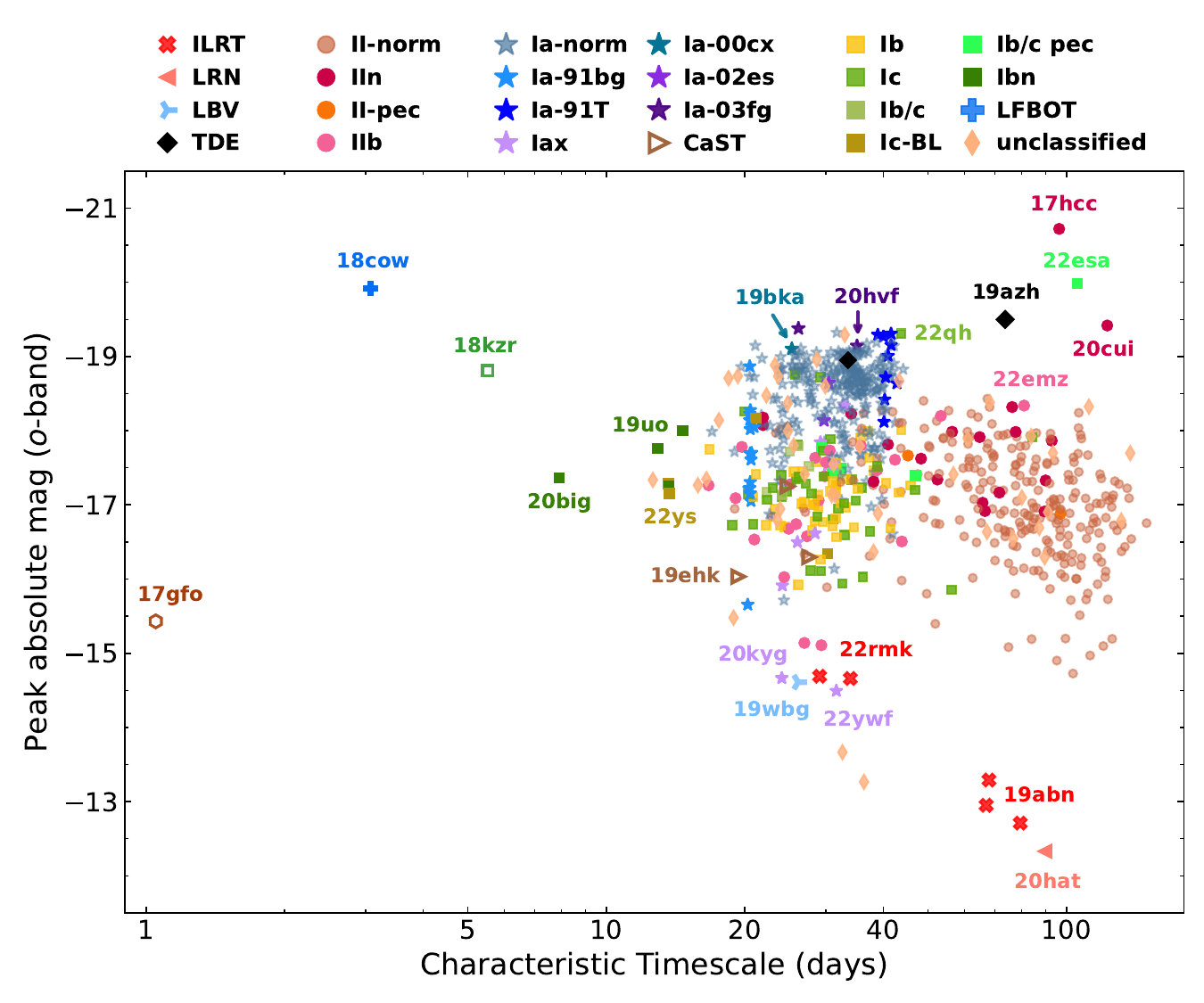}
\caption{Peak luminosity versus rest-frame timescale or duration in $o$-band for all transients in the ATLAS100 sample that passed the light curve quality cuts. A selection of individual events across different spectral types are highlighted in the plot. The rapidly evolving transients \ATxx{2017gfo} and \ATxx{2018kzr} are not part of the sample but are shown for context.}
\label{fig:luminosity_timescale}
\end{figure*}

A duration-luminosity diagram is a succinct way to visualise and compare different transient subclasses \citep{
2007Natur.447..458K,
2012PASA...29..482K, 2017ApJ...849...70V}. Figure~\ref{fig:luminosity_timescale} represents the ATLAS100 sample in this phase space, where the characteristic timescale or duration of the light curve and peak luminosity was estimated using the methodology described in Section~\ref{subsec:lcfit}. For context, the rapidly evolving kilonova \ATxx{2017gfo} \citep[e.g.][]{2017Natur.551...67P,2017Natur.551...75S} was added to the plot, along with other fast transients such as \ATxx{2018cow} \citep{2018ApJ...865L...3P,2019MNRAS.484.1031P} and \ATxx{2018kzr} \citep{2019ApJ...885L..23M,2020MNRAS.497..246G}. For \ATxx{2017gfo}, the light curve duration and $o$-band peak luminosity was estimated using a synthetic $o$-band light curve model generated following \cite{2021MNRAS.505.3016N}. AT\,2018cow is part of the ATLAS100 sample but did not pass the light curve quality cuts due to the rapid evolution. For \ATxx{2018cow} and \ATxx{2018kzr}, we use the duration and luminosity estimates provided by \cite{2020ApJ...904...35P}.

The ATLAS100 sample represents the most nearby and therefore most well-studied SNe, and there are several publications in the literature presenting detailed studies of individual events in the sample. ATLAS photometry has contributed extensively to several of these studies. Here we present a brief and non-exhaustive overview of the variety of transients in the ATLAS100 sample.

\subsubsection{Hydrogen-rich SNe}\label{subsubsec:SNII}

SNe II comprise the most abundant spectroscopic class within the ATLAS100 sample (Figure~\ref{fig:PieChart}). For the duration-luminosity diagram, we further subdivide SNe II into II-norm, IIn, II-pec and IIb (SNe IIb are discussed further in Section~\ref{subsubsec:SESN}). From the duration-luminosity diagram (Figure~\ref{fig:luminosity_timescale}), it is clear that SNe II-norm tend to have longer timescales, albeit with a large variation in both timescale and peak luminosity. The median timescale for II-norm that passed the light curve quality cuts
(251 out of the total 643) is 81.5 (27.6) days, where the standard deviation is indicated in the parentheses. The median peak absolute magnitude for SNe II-norm in the $o$-band is $-17.0$ (0.7) mag AB. SNe IIn on the other hand show a slightly lower median timescale of 65.7 (26.3) days, with a generally higher peak luminosity of $-17.9$ (0.9). 

The sample includes \SNxx{2023ixf}, one of the nearest and most well-studied SNe II in the last decade, with evidence of complex, asymmetric CSM \citep[e.g.][]{2023ApJ...956L...5B,2023ApJ...954L..42J}. The variable, red supergiant (RSG) progenitor of \SNxx{2023ixf} likely suffered multiple eruptive mass-loss episodes shortly before explosion \citep{2023ApJ...955L...8H,2023ApJ...952L..30J,2023ApJ...952L..23K}. Another nearby type II \SNxx{2024ggi} at $\sim 7$ Mpc was discovered by ATLAS within hours of explosion \citep{2024TNSAN.100....1S,2025ApJ...983...86C}, also showing evidence of CSM interaction \citep{2024ApJ...972L..15S} and a red supergiant progenitor identified in archival \textit{HST} and \textit{Spitzer} data \citep{2024ApJ...969L..15X}. \SNxx{2024ggi} is not part of the current sample but will be included in a future extension of the catalogue. The sample includes short plateau events such as \SNxx{2018gj} \citep{2023ApJ...954..155T} and \SNxx{2020jfo} \citep{2021A&A...655A.105S}, and low luminosity events like \SNxx{2018hwm} \citep{2021MNRAS.501.1059R}, \SNxx{2018is} \citep{2025A&A...694A.260D}, \SNxx{2018lab} \citep{2023ApJ...945..107P}, \SNxx{2020cxd} \citep{2021A&A...655A..90Y,2022MNRAS.514.4173K} and \SNxx{2021aai} \citep{2022MNRAS.513.4983V}. A sample study of 330 nearby SNe IIP within 200 Mpc in ZTF found 16 low luminosity ($M_r \geq -16$) events \citep{2025arXiv250620068D}, likely originating from low mass RSG progenitors. \SNxx{2020tlf} represents the first case of a precursor detected in a normal SN IIP/L \citep{2022ApJ...924...15J}, starting months prior to explosion.


Interacting SNe IIn often show precursor outbursts months to years preceding terminal explosion \citep{2014ApJ...789..104O,Smith2016}, with $\sim 26\%$ of ZTF SNe IIn showing a precursor brighter than $-13$ within three months of explosion \citep{2021ApJ...907...99S}. Several SNe IIn in the ATLAS100 sample show precursor outbursts detected in pre-explosion imagery -- these include \SNxx{2018cnf} \citep{2019A&A...628A..93P}, \SNxx{2020pvb} \citep{2024A&A...686A..13E} and \SNxx{2023ldh} \citep{2025arXiv250323123P}. Owing to its sensitivity and long baseline, \PS\  \citep{2016arXiv161205560C} is well-suited for detection of faint precursors as shown in the search for faint transients in the Pan-STARRS surveys by
\cite{2025MNRAS.542..541F}. 

A forthcoming study will focus on the subsample of SNe IIn within 100 Mpc with ATLAS, combined with archival \PS\ photometry to look for precursor outbursts. We note here that out of the 41 SNe IIn in this sample with declination $\delta > -40$\degr, 7 events ($17\%$) have precursors in \PS. This fraction is not corrected for survey sensitivity, cadence, sky coverage etc. and therefore represents a lower limit; the true fraction is presumably higher \citep{2024A&A...686A.231R}.

\subsubsection{Stripped Envelope SNe}\label{subsubsec:SESN}

The progenitors of SESNe (IIb, Ib and Ic)  have been proposed to be more massive main-sequence stars than their H-rich CCSN counterparts \citep{2011MNRAS.414.2985D}, 
although mass transfer in close binary systems likely plays a significant role in stripping the envelope \citep{1992ApJ...391..246P,2014ARA&A..52..487S,2017ApJ...840...10Y}. Explosion parameters and ejecta masses estimated for large samples of SESNe are generally not as extreme as one would expect if single, massive Wolf-Rayet (WR) progenitors were the dominant channel, favouring a binary progenitor for most SESNe \citep[e.g.][]{2013MNRAS.434.1098C,2016MNRAS.457..328L,2018A&A...609A.136T,2019MNRAS.485.1559P}.

SNe Ib and Ic in the ATLAS100 sample occupy broadly similar regions in the duration-luminosity diagram, with median peak absolute magnitudes of $-17.1\,(0.4)$ and $-17.2\,(0.8)$, and median timescales of $30.9\,(6.2)$ days and $29.3\,(11.6)$ days for Ib and Ic respectively. SNe IIb show similar median peak luminosity of $-17.3\,(0.8)$ and timescale of $28.7\,(13.7)$ days, but with a larger scatter in their light curve duration. In our sample, SNe Ic-BL appear to have a shorter median duration than both the narrower lined Ib, Ic and Ib/c events. While we have only 4 objects that meet the $N_{\rm cut}$ criterion, they do not appear significantly brighter or broader than the rest of the Ibc sample 
(see Table\,\ref{tab:summary}), with a median duration of 17.4 (6.8) days, in contrast to $\sim 30$ days duration for the rest of the SESN population. This is also seen in the sample of  \cite{2019MNRAS.485.1559P} at lower statistical significance. It is perhaps surprising that SNe Ic-BL show shorter durations compared to the rest of the SESN population, but previous work has combined inhomogeneous samples of SNe Ic-BL over large redshift ranges. We defer a quantitative analysis of the ejecta and $^{56}$Ni masses to a future paper, in order to constrain powering mechanisms \cite[as in][]{2024Natur.628..733R}. 

The sample includes the extraordinary Ic \SNxx{2022jli} that exhibited a $\sim 12.5$ day periodicity in its light curve \citep{2023ApJ...956L..31M,2024Natur.625..253C,2024arXiv241021381C}, likely originating from the newly formed compact object (neutron star or black hole) interacting with its binary companion. The unusual and luminous Ic event \SNxx{2022esa} exhibits a similar periodicity of $\sim 32\,$d with multiple cycles over 200\,d \citep{2025PASJ..tmp..134M}. \SNxx{2022esa} is classified on the TNS as SN Ia-CSM; we reclassify it as SN Ic-pec in our sample. \SNxx{2023aew} was another enigmatic transient with a double-peaked light curve, transforming from a SN IIb during the initial peak to a SN Ib/c during the second peak \citep{2024A&A...689A.182K,2024ApJ...966..199S}. The Ic-BL \SNxx{2022xxf} also showed a double-peaked light curve \citep{2023A&A...678A.209K}, where the second peak was interpreted as CSM interaction with H/He-poor material, likely expelled from the progenitor shortly before explosion. The Ib SNe 2019oys and 2019yvr showed evidence of interaction with H-rich CSM at late times \citep{2020A&A...643A..79S,2024MNRAS.529L..33F}, whereas the interacting Ibn \SNxx{2023fyq} displayed precursor activity from months prior to explosion \citep{2024A&A...684L..18B,2024ApJ...977..254D}. The sample also includes the Ibn \SNxx{2019uo} \citep{2020ApJ...889..170G} and the Icn \SNxx{2019jc} \citep{2022ApJ...938...73P}.

A rare and rather elusive subpopulation of long-duration SESNe was recently reported by \cite{2023A&A...678A..87K} in PTF survey data, accounting for $\sim 6\%$ of the SESN rate. These long-duration SESNe show a preference for low-metallicity, star-forming hosts, favouring massive progenitors with much higher ejecta masses compared to normal Ib/c events \citep{2023A&A...678A..87K}. Additionally, \cite{2022ApJ...941..107G} studied a population of luminous SNe or LSNe with $-19 > M_r^{\rm peak} > -20$. These events form a link between normal SNe Ib/c and SLSN-I populations in terms of their durations and peak luminosities, and could be powered by an additional energy source such as magnetar spin-down along with radioactive $^{56}$Ni \citep{2022ApJ...941..107G}.

We identify a handful of SESNe in the duration-luminosity diagram with unusually long timescales of $\gtrsim 40$ days. These include the luminous Ic \SNxx{2022qh} with $M_o^{\rm peak} \approx -19.3$, and the surprisingly long-lived Ic \SNxx{2020sgf} and IIb \SNxx{2022emz} with timescales of $\gtrsim 80$ days. A detailed investigation of SESNe in ATLAS, with particular emphasis on these long-rising and long-duration SESNe will be presented in a forthcoming paper. 


\subsubsection{Thermonuclear SNe}\label{subsubsec:SNIa}

SNe Ia-norm form a reasonably tight cluster in the duration-luminosity diagram, although there are a few highly extinguished events due to host reddening at the low luminosity end. The median peak absolute magnitude of $-18.7\,(0.6)$ and median timescale of $33.4\,(5.6)$ days reflects the relative homogeneity as expected. The Ia-91T and Ia-91bg sub-classes have median peak absolute magnitudes of $-19.0\,(0.4)$ and $-17.7\,(0.7)$, with median timescales of $40.5\,(1.1)$ days and $20.6\,(0.2)$ days, respectively. SNe Iax are the most diverse subclass with a median absolute magnitude of $-16.6\,(1.3)$. 
Despite their relative homogeneity, a rich diversity within thermonuclear SNe is undeniably present \citep{Taubenberger2017}, hinting at multiple progenitors and/or explosion mechanisms.

We highlight the interesting event \SNxx{2019bka} classified as Ia-91T on the TNS \citep{2019TNSCR.317....1S}. Although the classification spectrum does resemble Ia-91T, we measure a duration of only $\sim 24$ days, significantly faster evolving than the rest of the Ia-91T sub-population in the sample (with median dutation of $40.5 \pm 1.1$ days). \SNxx{2019bka} thus stands out in the duration-luminosity diagram as an outlier among Ia-91T events. The transient is in a remote location, with a projected separation of $41\,$kpc from the elliptical galaxy CGCG 097-109. Given these properties, \SNxx{2019bka} possibly belongs to the rare subclass of peculiar 00cx-like SNe Ia \citep{2013MNRAS.436.1225S}. 

The faintest SNe Iax in the sample are \SNxx{2019gsc} \citep[$M_r^{\rm peak} \approx -14.0$;][]{2020ApJ...892L..24S,2020MNRAS.496.1132T}, \SNxx{2022ywf} ($M_o^{\rm peak} \approx -14.5$) and \SNxx{2020kyg} \citep[$M_o^{\rm peak} \approx -14.7$;][]{2022MNRAS.511.2708S}. \SNxx{2019muj} \citep[$M_r^{\rm peak} \approx -16.4$;][]{2021MNRAS.501.1078B} and \SNxx{2019ovu} ($M_o^{\rm peak} \approx -16.5$) are intermediate luminosity Iax events, whereas \SNxx{2020sck} \citep[$M_o^{\rm peak} \approx -17.8$;][]{2022ApJ...925..217D} and \SNxx{2020udy} \citep[$M_o^{\rm peak} \approx -18.3$;][]{2023MNRAS.525.1210M,2024ApJ...965...73S} represent the luminous end of the distribution. The faint Iax events \SNxx{2019ttf} ($M_r \approx -13.7$, although with likely significant host extinction), \SNxx{2021fcg} \citep[$M_r^{\rm peak} \lesssim -13$;][]{2021ApJ...921L...6K} and \SNxx{2023bsr} ($M_o^{\rm peak} \approx -13.5$) were discovered during the time window of this sample, but were not registered to TNS since they did not generate at least three $5\sigma$ detections in ATLAS on any single epoch, although forced photometry does reveal lower significance ($3\sigma$) detections around peak for all three events. 

The sample includes well-studied Ia-norm events with an early excess in their light curves such as \SNxx{2018oh} \citep{2019ApJ...870L...1D,2019ApJ...870...12L,2019ApJ...870...13S}, \SNxx{2019np} \citep{2022MNRAS.514.3541S}, \SNxx{2021hpr} \citep{2023ApJ...949...33L} and \SNxx{2023bee} \citep{2023ApJ...953L..15H,2024ApJ...962...17W}. This feature has been attributed to different powering mechanisms such as ejecta-companion interaction \citep{2010ApJ...708.1025K}, ejecta-CSM interaction \citep{2016ApJ...826...96P}, surface $^{56}$Ni distribution \citep{2020A&A...642A.189M}, and helium shell detonation \citep{2019ApJ...873...84P}.
We note however that the term ``early excess'' is quite broad and not straightforward to quantify, since it depends on the cadence and quality of the observed data and the model/method used to fit the early light curve. The morphology of the early excess features reported in the aforementioned Ia-norm events is markedly distinct from the prominent spike-like, non-monotonic early excess features \citep{2018ApJ...865..149J,2024ApJ...966..139H} seen in recent Ia-02es and Ia-03fg events like iPTF14atg \citep{2015Natur.521..328C}, \SNxx{2019yvq} \citep{2020ApJ...898...56M,2021ApJ...919..142B}, \SNxx{2020hvf} \citep{2021ApJ...923L...8J}, \SNxx{2021zny} \citep{2023MNRAS.521.1162D}, \SNxx{2022ilv} \citep{2023ApJ...943L..20S}, \SNxx{2022vqz} \citep{2024MNRAS.527.9957X}, \SNxx{2022ywc} \citep{2023ApJ...956L..34S}, \SNxx{2022abom} \citep{2025MNRAS.542..541F} and \SNxx{2021qvo} \citep{2026ApJ...997..261P}. Among these events, SNe 2019yvq, 2020hvf and 2022vqz are part of the ATLAS100 sample. Interaction with $\sim 10^{-3} - 10^{-1}$ \msun\ of CSM placed $\sim 10^{13} - 10^{15}$ cm from the WD progenitor is a promising mechanism to explain the duration, colour and luminosity of these observed early excess features \citep[e.g.][]{2016ApJ...826...96P,2023MNRAS.522.6035M}. The CSM itself could plausibly arise as a natural outcome of mass loss in a double-degenerate system as tidally stripped material from the secondary WD \citep{2013ApJ...772....1R,2014MNRAS.438...14D}, or from wind-driven material launched in polar directions from the accretion disk formed following the disruption of the secondary WD \citep[disk-originated matter;][]{2017MNRAS.470.2510L}. Nebular spectra of some members belonging to both Ia-02es and Ia-03fg subclasses have shown narrow [\ion{O}{i}] emission \citep{2013ApJ...775L..43T,2016MNRAS.459.4428K,2019MNRAS.488.5473T,2023MNRAS.521.1162D,2024ApJ...960...88S}, considered to be a signature of a violent WD merger \citep{2012ApJ...747L..10P,2013ApJ...778L..18K}. Moreover, mid-infrared spectra from \textit{JWST} revealed strong [\ion{Ne}{ii}] emission at $12.81\,{\rm \mu m}$ in the nebular spectra of the Ia-03fg \SNxx{2022pul} \citep{2024ApJ...966..135K}, a feature that is only predicted by models involving a violent merger \citep{2023A&A...678A.170B}. This highlights the diagnostic power of combining early-time and late-time observations, both of which favour a WD merger scenario for these peculiar SN Ia subtypes. A sample of recent Ia-02es and Ia-03fg events displaying early excess features in combined ATLAS, ZTF and \PS\ data will be presented in a forthcoming paper.

Beyond their utility for in-depth studies of individual objects and subpopulations, low-redshift SNe Ia play a critical role in anchoring the Hubble diagram for cosmological analyses \citep{2018MNRAS.475..193F,2025MNRAS.541.2585V}. 
With the advent of next-generation surveys like the Legacy Survey of Space and Time \citep[LSST;][]{2019ApJ...873..111I}, which will observe unprecedented numbers of high-redshift SNe Ia, the precision of cosmological constraints will increasingly rely on the quality and completeness of the low-$z$ anchor sample. A well-characterised nearby SN Ia population with uniform selection and well-understood systematics is essential to control calibration uncertainties and standardisation biases that propagate through to measurements of dark energy parameters. The ATLAS survey, with its high cadence, wide field of view and all-sky coverage, and 10-year catalogue, is not only well-suited to providing a cosmologically useful low-$z$ SN Ia sample, but is actively being used to prepare such a sample. A coordinated series of forthcoming papers will present the first data release of the cosmological ATLAS SN Ia sample, including details of photometric calibration \citep{2025arXiv251221903M}, light curve cleaning and fitting, and survey simulations.

\subsubsection{Other Transients}

A small fraction ($2.3\%$) of transients in ATLAS100 is classified as ``Other'' as discussed in Section~\ref{subsubsec:SNtypes} (Figures~\ref{fig:PieChart} and \ref{fig:PieChartOther}). The Other category encompasses an eclectic and interesting mix of transient classes including LFBOTS, TDEs, LBVs, LRNe, ILRTs, and CaSTs. 

The ATLAS100 sample includes four known and well-studied TDEs -- \ATxx{2019azh} \citep[e.g.][]{2022ApJ...925...67L}, \ATxx{2019qiz} \citep[e.g.][]{2020MNRAS.499..482N,2021ApJ...917....9H}, \ATxx{2021ehb} \citep{2022ApJ...937....8Y} and \ATxx{2022dsb} \citep{2024MNRAS.531.1256M}. 

The sample includes well-studied LRNe like \ATxx{2017jfs} \citep{2019A&A...625L...8P}, \ATxx{2018bwo} \citep{2021A&A...653A.134B}, \ATxx{2020hat} and \ATxx{2020kog} \citep{2021A&A...647A..93P}, \ATxx{2021afy} and \ATxx{2021blu} \citep{2023A&A...671A.158P} and \ATxx{2021biy} \citep{2022A&A...667A...4C}. A massive progenitor candidate of $\sim 50$ \msun\ was identified in archival data for the luminous LRN \ATxx{2021aess} (Guidolin et al. 2026, A\&A, submitted), also part of the sample. The volumetric rate of LRNe is a strong function of luminosity \citep{2023ApJ...948..137K}, with luminous LRNe such as \ATxx{2021aess} being intrinsically rare. LRNe are generally considered to be the outcome of non-terminal events associated either with common envelope ejections or stellar mergers in close binary systems \citep{2013Sci...339..433I,2017ApJ...834..107B}. ILRTs on the other hand may be terminal electron capture SNe (ECSNe) from super-AGB progenitors \citep{2009MNRAS.398.1041B}. ILRTs in the sample include \ATxx{2018aes} \citep{2021A&A...654A.157C}, \ATxx{2019abn}, \ATxx{2019ahd} and \ATxx{2019udc} \citep{2025A&A...695A..42V,2025A&A...695A..43V}, and \ATxx{2022fnm} \citep{2024A&A...688A.161M}. 

LBVs comprise another diverse subclass of gap transients associated with non-terminal eruptions in evolved, massive stellar systems \citep{2011MNRAS.415..773S}. Also referred to sometimes as ``SN-impostors'', LBV/impostor events have been observed as precursors of interacting SNe in a handful of cases \citep[e.g.][]{2007Natur.447..829P,2013ApJ...779L...8F}. 
Owing to their intermediate luminosity and erratic, often rapid decline rates, luminous LBV eruptions occupy a region overlapping with kilonovae in the duration-luminosity diagram and were identified as major contaminants in the search for kilonovae in Pan-STARRS survey data \citep{2021MNRAS.500.4213M,2025MNRAS.542..541F}. Their bluer colours and longer survey baselines (capturing multiple eruptions) can be used to distinguish them from kilonovae.

CaSTs constitute a family of low-luminosity transients with poorly understood progenitors, with recent evidence for distinct subpopulations \citep[Ca-Ia and Ca-Ib/c;][]{2020ApJ...905...58D}, indicating multiple underlying progenitor channels could be responsible. These include core-collapse in highly stripped stars \citep{2017ApJ...846...50M} and weak explosions involving low-mass WD progenitors \citep{2017ApJ...836...60L}. A significant fraction of CaSTs occur at large projected separations from early-type host galaxies \citep{2014MNRAS.444.2157L}. The host environments for this Ca-Ia subpopulation show no evidence of current star formation and favour old stellar populations, interpreted as evidence for WD progenitors. Interestingly, a number of recent, nearby CaSTs discovered young show double-peaked light curves -- these include \SNxx{2019ehk} \citep{2020ApJ...898..166J}, \SNxx{2021gno} \citep{2023MNRAS.526..279E}, \SNxx{2021inl} \citep{2022ApJ...932...58J} and \SNxx{2022oqm} \citep{2024ApJ...962..109I,2024ApJ...972..194Y}, all part of the ATLAS100 sample. The primary peak (or early excess) in \SNxx{2021gno} and \SNxx{2021inl} that lasts for a few days was modelled using interaction with confined CSM \citep{2022ApJ...932...58J}, who inferred a CSM mass of $\lesssim 10^{-2}$ \msun\ and a radial extent of $\sim 10^{13}-10^{14}$ cm. Moreover, luminous and rapidly decaying X-ray emission within a day of explosion has been detected for two CaSTs, \SNxx{2019ehk} \citep{2020ApJ...898..166J} and \SNxx{2021gno} \citep{2022ApJ...932...58J}, consistent with the CSM parameters derived from modelling the multi-colour light curve during the early excess \citep{2022ApJ...932...58J}. We note that these CSM parameters are similar to those inferred from modelling the early excess features observed in recent Ia-02es and Ia-03fg events (Section~\ref{subsubsec:SNIa}), suggesting WD mergers may be the unifying link across a range of peculiar thermonuclear SN subclasses spanning faint CaSTs, to subluminous Ia-02es and over-luminous Ia-03fg events. Variations in primary WD mass, binary mass ratio, WD compositions and the inherently asymmetric nature of WD mergers could potentially provide a natural explanation for the diversity in observed properties across these subtypes. Recent simulations of low-mass WD mergers, involving either a hybrid HeCO WD disrupting a CO WD \citep{2023ApJ...944...22Z}, or a CO WD merging with a He WD \citep{2024A&A...683A..44M,2025arXiv250312105C} predict weak/faint thermonuclear explosions and thus present a promising avenue for further investigation.


\subsection{Sample purity and completeness}\label{subsec:compl}



In this section, we attempt to quantify the purity and completeness of the ATLAS100 sample. 

\subsubsection{Redshift completeness of galaxy catalogues and implications for completeness}
We have used a combination of host galaxy redshifts and redshifts derived from the transient (or its host) during follow-up. Here we consider our completeness due to both of these selection effects. 

\begin{figure}
\includegraphics[width=\linewidth]{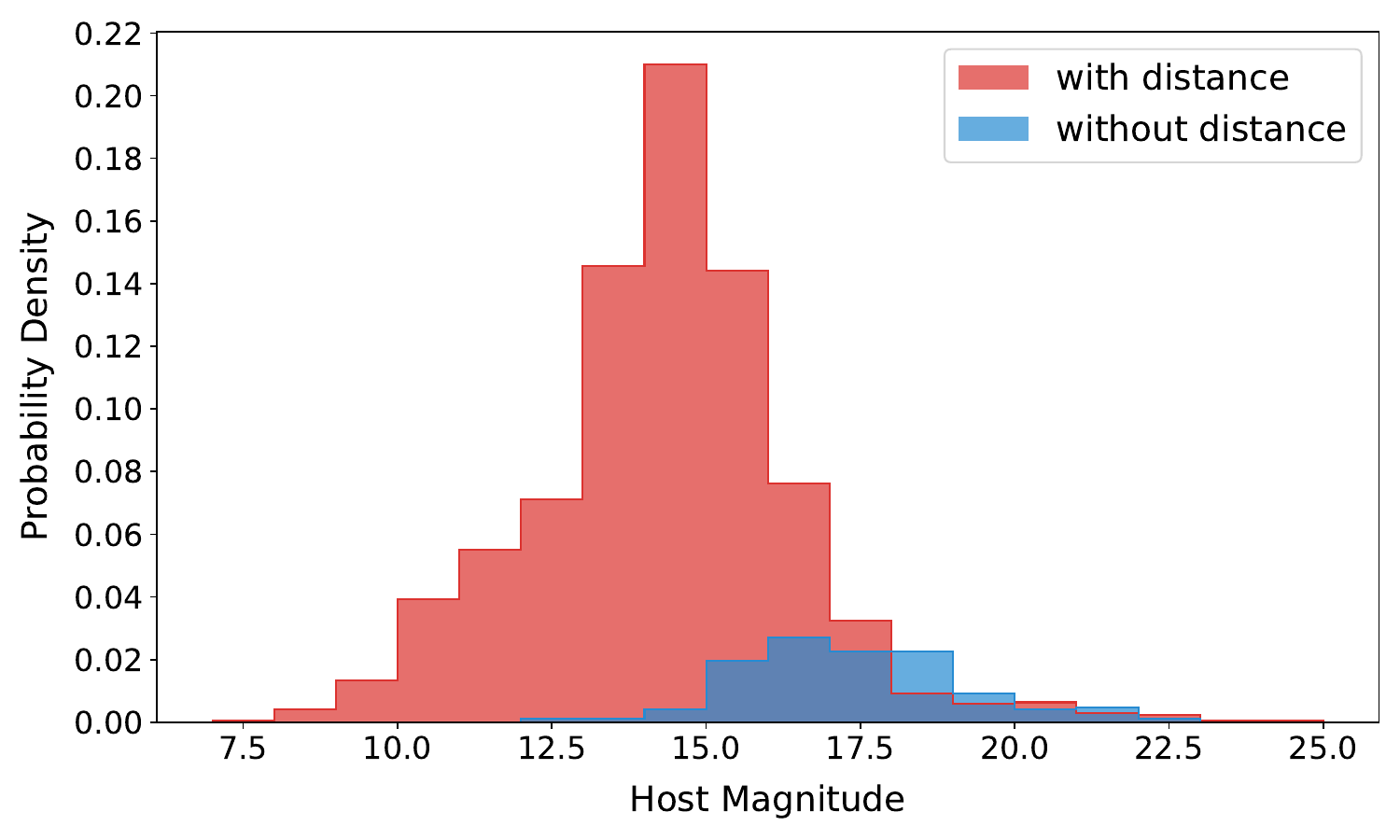}
\caption{Histogram showing the brightness distribution of host galaxies in ATLAS100 from \sherlock. A majority (83\%) of the host galaxies have a catalogued redshift (shown in red). The hosts without a prior catalogued redshift (shown in blue), where the transient redshift was inferred from the classification spectrum, represent a fainter population of host galaxies.}
\label{fig:sherlock_mags}
\end{figure}

\begin{figure}
\includegraphics[width=\linewidth]{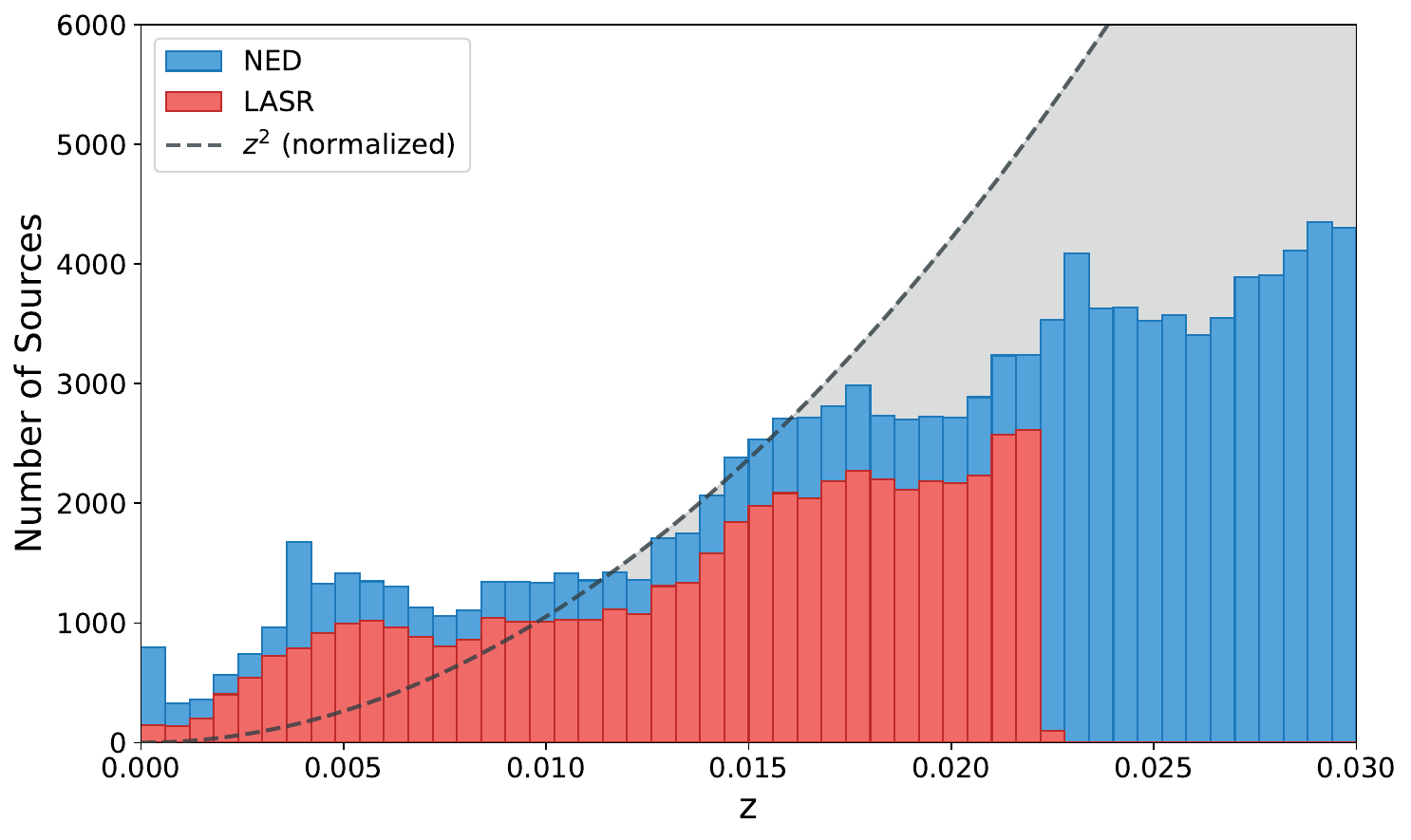}
\caption{Histogram showing the redshift distribution of sources in the LASr \citep{2020MNRAS.494.1784A} and NED-LVS \citep{2023ApJS..268...14C} catalogues within a redshift of 0.03. Since each redshift bin contains differential galaxy counts, we plot $z^2$ (dashed line) to represent the volume}.
$z^2$ is scaled to match the counts in the redshift bin at $z=0.01$, since the local galaxy distribution at low redshift is dominated by the Local Group and Virgo Supercluster.
\label{fig:sherlock_NED_LASr}
\end{figure}

The redshift completeness fraction, or RCF, is defined as the fraction of SN host galaxies with known spectroscopic redshift information (prior to the discovery of the transient) relative to the total number of transients in the sample. Every transient in the sample has a securely identified  host galaxy detected in optical imaging. We find no case of a transient which has a redshift within the $z\leq0.025$ limit that has either no detection of a nearby host or is separated by a large spatial extent ($R_{\rm proj}\geq50$\,kpc) and has a discrepant redshift.  
Essentially, almost every transient is associated with a unique host that can be optically identified. In a handful of rare cases (\SNxx{2019be}, \SNxx{2019bkc} and \SNxx{2020ags)}, the location of the transient is within a cluster of galaxies where a unique host association is not obvious. However, the redshift of the transient inferred from the classification spectrum is consistent with the spectroscopic redshift of the potential hosts, and is thus secure.
Therefore the RCF is a measure of the redshift completeness  down to a galaxy brightness, or mass, limit. The RCF is expected to be fairly high in the local volume, falling off steeply as a function of distance and host galaxy luminosity \citep{2020ApJ...895...32F}. \cite{2018ApJ...860...22K} estimated an RCF of $\sim 80\%$ for $z \leq 0.03$, corresponding to a distance of $\sim 130\,$Mpc. The ATLAS100 sample has a RCF of $83\%$; i.e. prior host galaxy redshifts were not available for $17\%$ of the spectroscopically classified transients in the sample. This is consistent with previous expectations for RCF in the local volume \citep{2018ApJ...860...22K,2020ApJ...895...32F}. For this $17\%$, the redshifts come mostly from the broad lines of the SN; e.g. from the redshift derived through the cross-correlation of the spectrum in \textsc{snid}. We examine the distribution of host galaxy magnitudes with and without a prior catalogued redshift in Figure~\ref{fig:sherlock_mags}. It is clear that the hosts identified by \sherlock\ without a catalogued redshift represent a fainter population, as expected. Finally, we examine the redshift completeness of the NED Local Volume Sample \citep[NED-LVS;][]{2023ApJS..268...14C} and LASr \citep{2020MNRAS.494.1784A} catalogues, the two major sources for a host galaxy spectroscopic redshift for \sherlock\ within 100\,Mpc (Figure~\ref{fig:sherlock_cat}). NED-LVS is a subset of NED consisting of $\sim 2$ million galaxies within 1000\,Mpc. We cleaned the NED-LVS catalogue by removing photometric redshifts, and any redshifts marked with an ``unreliable'' qualifier. Figure~\ref{fig:sherlock_NED_LASr} shows the redshift distribution of galaxies in the NED-LVS and LASr catalogues within $z \leq 0.03$. The dashed line representing volume is scaled to the redshift bin at $z=0.01$, since at low redshift the distribution of galaxies is governed by the Local Group and Virgo Supercluster.

We also consider any other kinds of transients we may be missing within 100\,Mpc. The possibilities for missing transients are: 
\begin{itemize}
    \item  Very remote transients: unclassified transients with a projected separation higher than $50\,$kpc from their hosts (since if they were classified on the TNS, a redshift would be available). 
    
    \item Faint transients in faint hosts: unclassified transients in galaxies with no catalogued spectroscopic redshift. These hosts may either have been missed by spectroscopic programmes, or be too faint for multi-object surveys. 
    
    \item Classified transients with an error in the spectroscopic redshift of the host or the SN spectrum. The most likely are transients at the edge of the $z\leq0.025$ limit which have a small error that places them beyond the cut.
\end{itemize}


To further quantify the missing fraction of transients in ATLAS100, we looked for bright, unclassified transients in ZTF BTS, that are not part of the ATLAS100 sample. The BTS has a very high spectroscopic completeness ($97\%$) for bright transients with $m < 18\,$mag \citep{2020ApJ...904...35P}. We find 40 transients in BTS within the relevant time window during 2017--2023, with a peak magnitude brighter than 17.5\,mag, lacking a spectroscopic classification, and not already present in ATLAS100. The light curves and contextual information of these transients were examined, such as image stamps and host galaxy information to assess any genuine unclassified transients missing in our sample. We find five events that were classified as foreground CVs on the TNS, but the spectral type was not updated on BTS. 30 of these events show light curves and contextual information that is consistent with foreground CVs. We also find 3 nuclear transients that were likely AGN flaring activity. Finally, 2 events were known SNe within 100 Mpc and present in ATLAS100, but evaded the crossmatch due to duplicate TNS entries. Thus, the overwhelming majority of these transients are either confirmed or likely foreground CVs. We do not find any bright, unclassified transient in BTS missing in ATLAS100. The possibility of our sample missing intrinsically and apparently faint transients in galaxies that are also faint and lack redshift information remains. This should be addressed with LSST by Rubin and the 4MOST instrument that will measure galaxy redshifts to fainter limits than previously achieved combined with a systematic spectroscopic survey  \citep[TiDES;][]{2025ApJ...992..158F} of sources brighter than $m\lesssim22$. 


\subsubsection{ATLAS100 and ZTF BTS}

Most of the transients in our sample are equatorial or northern, with declination $\delta > -30^{\circ}$ (Figure~\ref{fig:SNdistribution}), given the later commissioning of the southern ATLAS units. We thus expect a significant overlap between the ATLAS100 and ZTF BTS samples. A crossmatch yields only 441 transients in ATLAS100 that are also present in BTS. The BTS started on 01 May 2018, slightly later than ATLAS100. Even after accounting for 194 transients with $\delta < -30^\circ$ (limit for ZTF) and 119 transients that were discovered before 01 May 2018 in ATLAS100, the overlap (441 out of 1416, or $31\%$) is still somewhat small. This is likely owing to the strict quality cuts on light curves imposed for BTS that remove nearly half their sample \citep{2020ApJ...904...35P}. To confirm, we checked the 975 transients in ATLAS100 that are missing in BTS, with discovery date after 01 May 2018 and $\delta > -30^\circ$. A majority (656 out of 975) were detected and registered to TNS by ZTF, but were excluded from the BTS sample due to the aforementioned quality cuts.

We also reviewed the demographics of the 975 transients missing from BTS, and whether they represent a similar or distinct population relative to the overall ATLAS100 sample. Of the 975 transients, $87\%$ are classified, identical to the overall sample. SNe II constitute $41\%$ and SNe Ia $31\%$ of the 975 missing objects, again identical to the overall ATLAS100 sample. This shows that the ATLAS100 transients missing from ZTF BTS constitute the same underlying population, and the BTS light curve quality cuts do not preferentially select for or against any transient type. ATLAS100 thus comprises a complementary dataset with many objects not included in ZTF BTS. Although we do make quality cuts for fitting the light curves, we do not exclude any robustly identified transient within a redshift of 0.025, presenting a complete sample valuable for further investigation such as volumetric rate computations and global host galaxy properties.

\subsubsection{Missed detections}\label{subsubsec:misses}
A list of all transients reported to the TNS within the sample definition window, with a redshift of $z \leq 0.025$ was extracted and crossmatched with the ATLAS100 sample. We find 251 transients on TNS that are missing from our sample. This is attributable to several different factors noted below.

\begin{enumerate}
    \item Southern transients. The southern ATLAS units in Chile and South Africa were commissioned in late 2021, and became fully operational only by early to mid 2022. The coverage in the south was quite patchy in the initial period following commissioning during early 2022. We thus missed southern transients with $\delta \leq -40\degr$ that occurred before mid 2022. Out of the 251 transients, 70 were missed owing to the lack of southern sky coverage.

    \item Faint transients. A fair number of faint transients, mostly discovered by ZTF and \PS, were observed but peaked below the detection limit of ATLAS. Since these transients did not produce at least three $5\sigma$ detections in the difference images on any single epoch, they were not flagged and registered to the TNS and are thus missing from our sample. These account for 75 of the missing 251 transients.

    \item Solar conjunction and gaps in coverage. Some missing transients are old SNe that emerged from solar conjunction, with declining light curves discovered well past peak. In other cases, the peak of the light curve was missed due to gaps in survey coverage, either due to weather or downtime. In both such events, the transients had faded below the detection limit of ATLAS by the time they were observable. These account for 23 of the missing transients.

    \item Human error. In a small number of cases (5), the human scanner was not convinced by the veracity of the detection when presented with the difference images and forced photometry flux on the ATLAS transient science server marshall \citep{2020PASP..132h5002S}. These cases often involved faint transients detected just above the detection threshold, and/or transients located in proximity to the host galaxy nucleus or a bright foreground star. Because of the uncertainty, the transients were not promoted to the TNS and were instead placed in a ``possible'' list awaiting further data for confirmation. In the remaining few cases the transient was not promoted due to human error or oversight.

    \item Genuine misses. Finally, we note a  significant number (66) of transients that were observed by ATLAS and were brighter than typical ATLAS limiting magnitudes, as evidenced by their light curves from other surveys and forced photometry light curves retrieved from the ATLAS forced photometry server. As such these transients should have been flagged and promoted to the TNS, but were rejected during the data processing. Some of these transients were quite bright at peak and prompted us to investigate why they were missed. We find that they were detected by the PSF fitting routines by the ATLAS pipeline running on the images \citep[producing the .ddc files as described in][]{2018PASP..130f4505T,2020PASP..132h5002S}.
    A high fraction of these transients were nuclear or close to the core of a bright galaxy, resulting in a low initial real-bogus (RB) score from the machine-learning algorithm \citep{2020PASP..132h5002S} when the transient was faint. The low RB score resulted in the transient being consigned to the ``garbage'' list, and periodic purging of this list meant the candidate detection was discarded although subsequent detections from the rising transient were clearly significant. In other cases, \sherlock\ mistakenly identified and labelled the transient as a variable star due to proximity with a foreground star. 
    As part of the new routine operations of the ATLAS VRA, we now cross-match the garbage list with TNS once a week to identify potential misses, and whether they are from automated systems or human scanners. Since the ATLAS latitudes are well covered by other surveys such as ZTF and GOTO, this will eliminate misses of most transients, baring multiple algorithmic failures across independent systems, which is unlikely. 

    The remaining 12 objects had been previously vetted (Section~\ref{sec:vetting}) and rejected from the sample. These have a TNS object $z <= 0.025$, but either the identified \sherlock\ host has a catalogued redshift just over 0.025, or they are background impostors.
    
    
    
    
\end{enumerate}

\subsubsection{Low luminosity transients}

For a limiting magnitude of $m_o \approx 19$, the ATLAS100 sample is complete within 100 Mpc for transients brighter than an absolute magnitude, $M_o \approx -16$. The sample is not complete at 100 Mpc for transients with intrinsically low luminosity, such as some faint SNe II \citep{2024ApJ...969L..11D}, faint SNe Iax \citep{2022MNRAS.511.2708S} and several classes of gap transients like LRNe and ILRTs \citep{2023ApJ...948..137K}. In addition to the intrinsic luminosity of the transient, several other factors may also affect the recovery efficiency including foreground Galactic extinction, host extinction, variations in survey sensitivity or $5\sigma$ limiting magnitude, gaps in sky coverage due to weather or technical downtime, galactocentric distance and the element of human error. The galactocentric distance in particular seems to be an important factor as seen above in Section~\ref{subsubsec:misses}, since the higher level of noise in bright regions of the host galaxy affects the sensitivity, and image subtraction artefacts around bright galaxy cores also limit the discovery of fainter transients. 

A forthcoming paper on luminosity functions and volumetric rates will present recovery efficiency simulations for different classes of transients in the sample using the ATLAS survey simulation tool \citep{2021PhDOwen}. For a representative transient light curve as input, the code simulates a population of light curves distributed across a defined time window (taking into account survey history and sensitivity variations), sky coordinates (accounting for variations in Galactic extinction) and redshift within a defined volume, and performs an assessment of recovery \citep{2022MNRAS.511.2708S,2025MNRAS.542..541F}. 


\subsubsection{Sample purity}
\label{sec:purity}

The classified transient sample is very pure due to the careful, manual checks on both the host galaxy associations and the classification spectra. We find no obvious systematic source of contaminants in those with a classification spectrum and would consider the classified sample as 100$\%$ pure (in terms of being real transients). However, the classifications may be refined in the future with quantitative joint light curve and spectral analysis, so the classifications are not definitive even within the limitations of the qualitative classification scheme. 

It is possible the unclassified subsample may have contaminants, although in 
Figures\,\ref{fig:FirstMag} and \ref{fig:AngularSep} the distributions of these compared to the classified SN sample are indistinguishable. There 
may be some in this sample which are background SN contaminants, lying beyond the 100\,Mpc host but in chance alignment 
within the 50 kpc crossmatch radius projected on to an angular radius. A background normal SN Ia at $z \approx 0.1$ will peak at $m_o \approx 19$ mag, mimicking an intrinsically faint transient within 100 Mpc. In Section~\ref{sec:vetting}, we identified $\sim 60$ background contaminants during the vetting process. It is possible that a small number ($\lesssim 10$) of faint, unclassified background transients persist in our final sample.



\section{Conclusions}

This paper presents the ATLAS 100\,Mpc local volume survey (ATLAS100) sample of extragalactic transients within a redshift of $z \leq 0.025$ detected by the ATLAS survey during a span of 5.75 years from 2017 September 21 to 2023 June 21. The sample definition and careful vetting process is described in detail, resulting in a total of 1729 transients, with hydrogen-rich SNe II and SNe Ia constituting the bulk ($\sim 71\%$) of the sample. The spectroscopic classification completeness is good, with 87\% of transients having at least one optical classification spectrum. We find a redshift completeness fraction for galaxies of 83\% within this volume. The sample is very pure, close to 100\% for spectroscopically classified supernovae and we estimate a very low contamination rate for the unclassified transients. The completeness depends on the absolute luminosities of the various transients and will be distance dependent, which will be addressed in future work. The light curves of all the transients that passed certain quality cuts were fit to extract peak luminosity and characteristic timescale, defined as the time above half peak flux. We explore the statistical properties of the sample, including the demographics, host galaxy separations, and their location in the duration-luminosity diagram as a function of spectral type. This paper serves as a sample definition paper and marks the release of the catalogue and all light curve data, cleaned and binned into nightly data points.

\section*{Acknowledgements}

We thank the anonymous referee for constructive comments that improved the quality of this paper.

This work has made use of data from the Asteroid Terrestrial-impact Last Alert System (ATLAS) project. The Asteroid Terrestrial-impact Last Alert System (ATLAS) project is primarily funded to search for near earth asteroids through NASA grants NN12AR55G, 80NSSC18K0284, and 80NSSC18K1575; byproducts of the NEO search include images and catalogs from the survey area. This work was partially funded by Kepler/K2 grant J1944/80NSSC19K0112 and HST GO-15889, and STFC grants ST/T000198/1 and ST/S006109/1. The ATLAS science products have been made possible through the contributions of the University of Hawaii Institute for Astronomy, the Queen’s University Belfast, the Space Telescope Science Institute, the South African Astronomical Observatory, and The Millennium Institute of Astrophysics (MAS), Chile and the University of Oxford.  

This work is based (in part) on observations collected at the European Organisation for Astronomical Research in the Southern Hemisphere, Chile as part of PESSTO, (the Public ESO Spectroscopic Survey for Transient Objects Survey), ePESSTO and ePESSTO+ suveys. 

SJS, SS, KWS, HFS, FS, AJC and JHG acknowledge funding from the following grants held in Oxford:  STFC Grant ST/Y001605/1, a Royal Society Research Professorship and Newton International Fellowship, a Schmidt AI in Science fellowship and the  Hintze Family Charitable Foundation. 

AC has been supported by the ANID, through grant ICN12\_009 to the Millennium Institute of Astrophysics (MAS), and FONDECYT 1251692.
AP acknowledges support from the PRIN-INAF 2022 “Shedding light on the nature of gap transients: from the observations to the models”. PR acknowledges support from STFC grant 2742655. LR acknowledges support from the Trottier Space Institute Fellowship and from the Canada Excellence Research Chair in Transient Astrophysics (CERC-2022-00009). T.-W.C. acknowledges financial support from the Yushan Fellow Program of the Ministry of Education, Taiwan (MOE-111-YSFMS-0008-001-P1), and from the National Science and Technology Council, Taiwan (NSTC 114-2112-M-008-021-MY3).

MN, CRA, XS and AA are supported by the European Research Council (ERC) under the European Union’s Horizon 2020 research and innovation programme (grant agreement No.~948381).

\section*{Data Availability}

The binned and cleaned ATLAS light curves for all transients in the sample are publicly available online on the Oxford Research Archive (\url{https://ora.ox.ac.uk/objects/uuid:dbb60078-60ec-42af-8b50-53be5115de0a}). The methods used for processing the light curve data are described in Section~\ref{subsec:lcbin} and Appendix~\ref{app:offset}. The data release includes a catalogue file containing additional details about the transients, including redshifts, distances, updated classifications and host galaxy associations. 





\bibliographystyle{mnras}
\bibliography{references} 


\section*{Affiliations}
\noindent
{\it
$^{1}$Astrophysics, Department of Physics, University of Oxford, Denys Wilkinson Building, Keble Road, Oxford, OX1 3RH, UK\\
$^{2}$Astrophysics Research Centre, School of Mathematics and Physics, Queen's University Belfast, Belfast, BT7 1NN, UK\\
$^{3}$Space Telescope Science Institute, 3700 San Martin Dr, Baltimore, MD 21218, USA\\
$^{4}$Graduate Institute of Astronomy, National Central University, 300 Jhongda Road, 32001 Jhongli, Taiwan\\
$^{5}$INAF -- Osservatorio Astronomico di Padova, Vicolo dell'Osservatorio 5, I-35122, Padova, Italy\\
$^{6}$University Observatory, Faculty of Physics, Ludwig-Maximilians-Universität München, Scheinerstr. 1, 81679, Munich, Germany\\
$^{7}$European Southern Observatory, Alonso de C\'ordova 3107, Casilla 19, Santiago, Chile\\
$^{8}$Institute for Astronomy, University of Hawai'i, 2680 Woodlawn Drive, Honolulu, HI 96822, USA\\
$^{9}$Department of Physics and Astronomy, Johns Hopkins University, Baltimore, MD 21218, USA\\
$^{10}$Trottier Space Institute at McGill, 3550 Rue University, Montreal, Quebec H3A 2A7, Canada\\
$^{11}$Department of Physics, McGill University, 3600 Rue University, Montreal, Quebec H3A 2T8, Canada\\
$^{12}$GSI Helmholtzzentrum für Schwerionenforschung, Planckstraße 1, 64291, Darmstadt, Germany\\
$^{13}$Department of Astronomy and Steward Observatory, University of Arizona, 933 North Cherry Avenue, Tucson, AZ 85721-0065, USA\\
$^{14}$Instituto de Astrofísica, Pontificia Universidad Católica de Chile, Vicuña Mackenna 4860, Macul, Santiago, Chile\\
$^{15}$Instituto Milenio de Astrofísica (MAS), Nuncio Monseñor Sótero Sanz 100, Of. 104, Santiago, Chile\\
$^{16}$South African Astronomical Observatory, Cape Town, 7925, South Africa\\
$^{17}$Department of Physics, Stellenbosch University, Stellenbosch, 7602, South Africa\\
$^{18}$Instituto de Alta Investigaci\'on, Universidad de Tarapac\'a, Casilla 7D, Arica, Chile\\
$^{19}$School of Physics and Astronomy, University of Birmingham\\
$^{20}$Research School of Astronomy and Astrophysics, Australian National University, Canberra, ACT 0200, Australia\\
}

\appendix

\section{Baseline correction}
\label{app:offset}

For each transient, the forced photometry was measured from difference images (Section~\ref{subsec:lcbin}). Occasionally, the reference wallpaper used for image subtraction contains some flux from the transient while it is still bright, resulting in underestimation of the true transient flux. To correct the photometric measurements for such cases, we compute a baseline flux correction for each transient light curve in the sample generated by \texttt{ATClean}. This correction was computed using a pre-explosion window from $-220$ to $-20$ days relative to TNS discovery. The baseline correction was computed as a sigma-clipped median from the pre-explosion window. If the number of data points after sigma-clipping was insufficient for computing a robust correction ($N < 10$ for $o$-band and $N < 5$ for $c$-band), a late-time window from $+500$ to $+700$ days was used instead. The uncertainty on the correction was estimated using bootstrap resampling. 

The baseline correction was only applied if it was at least $3\sigma$ significance, and the uncertainty on the offset was below 50 $\mu$Jy. A significant baseline correction was computed and applied only to 275 light curves (246 unique transients) out of 3449 light curves (1729 unique transients). Most of these transients with a significant baseline correction were discovered during 2017--2019, when changes to the ATLAS reference wallpaper were more frequent. Overall, the applied corrections are generally small, with a median value of $-26$ $\mu$Jy. Although a majority of these cases show negative baseline correction values, indicating the presence of positive transient flux in the reference image, a small number (47 out of 275) of light curves have positive (albeit small) baseline corrections. We note that there could be other sources of additional flux besides changes to reference images, such as AGN flaring activity in the host galaxy for nuclear transients, or a spatially coincident foreground variable star. Although this is a small effect in an overall sense, for individual transients we encourage users to adopt a custom baseline correction if required, and we thus include both the corrected and uncorrected data products.

\begin{figure*}
\includegraphics[width=0.8\linewidth]{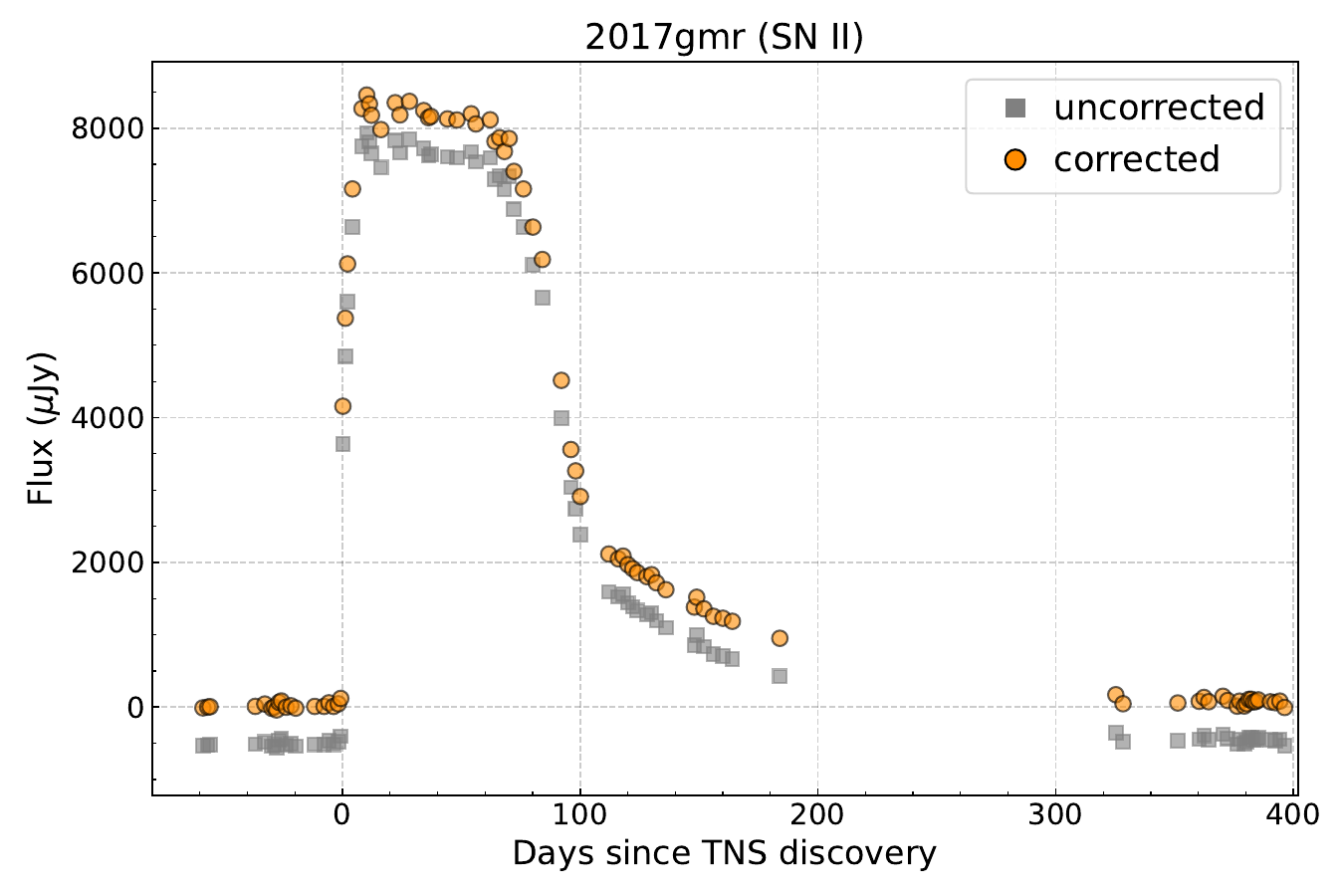}
\caption{ATLAS $o$-band light curve of the type II \SNxx{2017gmr}, shown with (orange circles) and without (grey squares) the baseline correction. A correction of $-524 \pm 9$ $\mu$Jy was computed for \SNxx{2017gmr} from the pre-explosion window before TNS discovery.}
\label{fig:offset_corr}
\end{figure*}

Figure~\ref{fig:offset_corr} shows the ATLAS $o$-band light curve of \SNxx{2017gmr} as an example, with and without the offset correction. The uncorrected light curve shows a constant negative offset, indicating presence of SN flux in the reference image.

\end{document}